\documentclass[acmsmall]{acmart}

\AtBeginDocument{%
  \providecommand\BibTeX{{%
    \normalfont B\kern-0.5em{\scshape i\kern-0.25em b}\kern-0.8em\TeX}}}

\setcopyright{acmcopyright}
\copyrightyear{2018}
\acmYear{2018}
\acmDOI{10.1145/1122445.1122456}

\acmJournal{POMACS}
\acmVolume{37}
\acmNumber{4}
\acmArticle{111}
\acmMonth{8}

\usepackage{caption}
\usepackage{subcaption}
\usepackage{booktabs}
\usepackage{graphicx}
\usepackage{multirow}
\usepackage{soul,color}

\definecolor{extreme}{HTML}{5E96AE}
\definecolor{bias}{HTML}{EE4540}
\definecolor{fake}{HTML}{32A853}
\definecolor{consp}{HTML}{00818A}
\begin{document}

%\title{Characterizing Roles and Information Flow in Extremist Information Mobilization on Facebook}
%\title{Educators, Solicitors, Flamers, Motivators, Sympathizers: Modeling Roles in Online Extremist Social Movements}
%\title{Educators, Solicitors, Flamers, Motivators, Sympathizers: Characterizing Roles and their Dynamics in Online Extremist Movements}
\begin{flushleft}
      {\color{red} To cite: Shruti Phadke and Tanushree Mitra. 2021. Educators, Solicitors, Flamers, Motivators, Sympathizers: Characterizing Roles in Online Extremist Movements. Proc. ACM Hum.-Comput. Interact. Computer Supported Cooperative Work (CSCW' 21), (accepted May 2021).}
      \vspace{10pt}
\end{flushleft}

\title{Educators, Solicitors, Flamers, Motivators, Sympathizers: Characterizing Roles in Online Extremist Movements}

\author{Shruti Phadke}
\affiliation{%
  \institution{Information School, University of Washington, USA}
  %\streetaddress{P.O. Box 1212}
  \city{Seattle, USA}
}
\author{Tanushree Mitra}
\affiliation{%
  \institution{Information School, University of Washington, USA}
  %\streetaddress{P.O. Box 1212}
  \city{Seattle, USA}
}

% \affiliation{%

% \affiliation{%

% \affiliation{%

% \affiliation{%

% \affiliation{%



\renewcommand{\shortauthors}{Anonymous Authors}

\begin{abstract}

      %The pervasiveness of online extremist ideologies, such as white supremacy and anti-LGBTQ, is a growing concern. 
      Social media provides the means by which extremist social movements, such as white supremacy and anti-LGBTQ, thrive online. %Be it spreading incendiary propaganda, putting out a call for action or educating readers about hateful ideologies, extremists increasingly use social media to coordinate and organize their extremist activism. 
      Yet, we know little about the roles played by the participants of such movements. In this paper, we investigate these participants to characterize their roles, their role dynamics, and their influence in spreading online extremism. Our participants---online extremists accounts---are 4,876 public Facebook pages or groups that have shared information from the websites of 289 Southern Poverty Law Center (SPLC) designated extremist groups. Guided by theories of participatory activism, we map the information sharing features of these extremists accounts. By clustering the quantitative features followed by qualitative expert validation, we identify five roles surrounding extremist activism---\textit{educators}, \textit{solicitors}, \textit{flamers}, \textit{motivators}, \textit{sympathizers}. For example, \textit{solicitors} use links from extremist websites to attract donations and participation in extremist issues, whereas \textit{flamers} share inflammatory extremist content inciting anger. We further investigate role dynamics such as, how stable these roles are over time and how likely will extremist accounts transition from one role into another. We find that roles core to the movement---\textit{educators} and \textit{solicitors}---are more stable, while \textit{flamers} and \textit{motivators} can transition to \textit{sympathizers} with high probability. 
      %We further investigate role dynamics such as,  how stable these roles are over time and how likely will extremist accounts transition fromone role into another. Finally, given the prevalence of mis- and disinformation in online extremist discourse, we also ask: How influential are the roles in spreading various types of information sources, such as extremist content, biased news, fake news, and conspiracies?  
      Finally, using a Hawkes process model, we test which roles are more influential in spreading various types of information. We  find that \textit{educators} and \textit{solicitors} exert the most influence in triggering extremist link posts, whereas \textit{flamers} are influential in triggering the spread of information from fake news sources.
      %Finally, using a robust temporal modeling technique, we find that \textit{educators} and \textit{solicitors} significantly influence the information sharing behavior of other roles. 
      Our results help in situating various roles on the trajectory of deeper engagement into the extremist movements and understanding the potential effect of various counter-extremism interventions. 
      Our findings have implications for understanding how online extremist movements flourish through participatory activism and how they gain a spectrum of allies for mobilizing extremism online. 
      %with the resources and affordances to spread extremism online.  
      
      %effective the online counter-extremism strategies can be in nudging the extremist accounts away from the online extremism. 
  
\end{abstract}




\maketitle

\section{\textbf{Introduction}}

 According to the Southern Poverty Law Center (SPLC)---a non profit organization dedicated to monitoring extremist activity in the United States---Facebook groups serve as the primary avenue for extremists to recruit new members and spread extremist propaganda \cite{Facebook74online}. Take for example, \textit{Pissed off White Americans}---a Facebook group harboring 64K followers on their public Facebook page. It describes itself as:  \textit{``We are the sons and daughters of European Heritage, and we are tired of being treated as second class citizens. LOUD, PROUD, and very PISSED OFF!.''} This page frequently posts links from a known white supremacy group, White Rabbit Radio---a Southern Poverty Law Center (SPLC) designated white nationalist group \cite{whiterabbit}. Their posts often contain calls for actions against white genocide---a white supremacy belief that there is a deliberate attempt to wipe out the white race to promote reproduction of other races.  %\textit{Pissed off White Americans} also urges their readers to like, share, and comment on the videos about white genocide, videos that are banned from YouTube due to violation of YouTube's hate speech policies. 
 Another Facebook group, \textit{White Lives Matter Movement} posts informational content explaining what it means to be white or ``aryan'' on their public Facebook page. Sharing links from the website of an SPLC designated neo-nazi group---National Vanguard \cite{Nationalvanguard}---they suggest norms for racial segregation and educate readers about white identity. While the \textit{Pissed off White Americans} page uses link sharing for \emph{soliciting} their readers' engagement, \textit{White Lives Matter Movement} Facebook page \emph{educates} their readers about white identity and cultural norms. In other words, both pages play different \textit{\textbf{roles}} in advancing their white supremacy ideology online. Moreover, by sharing links from the websites of known extremist organizations, both pages become participants in the extremist ecosystem on Facebook. We call them \textbf{\textit{extremist accounts}}---Facebook pages and groups involved in actively sharing information from any of the 289 Southern Poverty Law Center (SPLC) designated extremist groups, and thus, operating as key players in sustaining and growing the extremist movement.

%background
%In this paper, we focus on the U.S based domestic extremism such as white supremacy, anti-LGBTQ, anti-Immigration etc. Extremist movements are known to target the unsuspecting audience to promote their fundamentalist ideologies and agendas \cite{gerstenfeld2003hate,Levin2002}. Extremist movements vilify people  for the inherent characteristics such as race, nationality, religion, gender identity, sexual identity, and political views and \cite{Frequent30online} and often attempt to recruit new members who support the extremist cause \cite{Beirich2018}. Proponents of extremist ideologies have started utilizing affordances of online platforms to disseminate mis- and disinformation, calls for donations, invitations to alternative platforms with low moderation and links to podcasts and shows discussing extremist ideologies \cite{Crosspla63:online,Phadkecross,starbird2019disinformation}. 

In this work, we explore the ecosystem of such online \textit{extremist accounts} through the lens of participatory activism---the potential and magnitude of individuals and groups to engage in sociopolitical issues \cite{petrova2007transactional}. Previous scholars have largely studied the importance of participatory activism in positive social justice movements, such as feminist practices \cite{nunez2017online,keller2012virtual} or raising awareness against police brutality \cite{arda2015construction}. %and environmentalist movements \cite{petrovic2012public}. 
On the contrary, only a handful of studies have highlighted how anti-social movements, also adopt similar participatory activism to promote a positive notion of their brand \cite{krona20195}. For example, terrorist organizations such as ISIS used participatory activism practices to promote the caliphate as a way of life and recruited combatants \mbox{\cite{krona20195}}. Despite their clear presence on social media, studies analyzing online anti-social movements are rare. In particular, U.S based domestic extremist movements with white supremacy, anti-LGBTQ, or anti-Immigration agendas that predominantly operate online, have not been investigated through the lens of participatory activism. In our work, we bridge this gap. %and study participatory activism in online extremism. 
Using the lens of participatory activism, we consider extremist accounts to play various social roles towards advancing extremist movements online. Specifically, we identify 4,876 extremist accounts involved in sharing links from 289 Southern Poverty Law Center (SPLC) designated extremist groups using Facebook's CrowdTangle API. We obtain the public Facebook activity of 4,876 extremist accounts over two years (Jan'18 to Dec'19) to model their online roles in extremist movements. We further study their role dynamics and assess how influential roles are in spreading various types of information sources. We first ask:

\vspace{3pt}
\noindent \textbf{RQ1:  What social roles are played by extremist accounts in online extremist movements?}
\vspace{3pt}

% Extremist movements are known to target the unsuspecting audience to promote their fundamentalist ideologies and agendas \cite{gerstenfeld2003hate,Levin2002}. Their goals of vilifying people  for the inherent characteristics such as race, nationality, religion, gender identity, sexual identity, and political views often result in the recruitment of new members who support the extremist cause \cite{Beirich2018}. Examples in Figure \ref{fig:motivating_example} display that information can become a crucial resource catalyzing such extremist movements, especially in online setting. Proponents of extremist ideologies have started utilizing affordances of online platforms to disseminate mis- and disinformation, calls for donations, invitations to alternative platforms with low moderation and links to podcasts and shows discussing extremist ideologies \cite{Crosspla63:online}. Scholars have perceived this growing ecosystem of biased and polarized information as a threat to the digital democracy and a gateway to collective action \cite{starbird2019disinformation}. 

To answer, we consider extremism as a social movement---a collective effort by a group of people aimed towards changing the society in a way that aligns with the movement's goals \cite{diani1992concept}. Guided by theories of participation in social movements, we explore underlying behaviors of the accounts across three dimensions: drives for participation, engagement trends, and strategies of mobilization. Using features informed from these dimensions and qualitative expert validation, we identify five roles played by extremist accounts in forwarding their social movements:  \textit{solicitors}---who solicit participation and funds for the extremist movement, \textit{educators}---accounts that share intellectual content about extremism and prominently share and like extremist content, \textit{flamers}---accounts that express and incite anger by posting inflammatory content, \textit{motivators}---who are achievement oriented and go-getters of the extremist community and who post information that portrays a positive image of their extremist agenda, and \textit{sympathizers}---accounts that are fringe supporters of the extremist movement who sparingly engage with links from the extremist websites. 

These roles are based on the social media activities of the extremist accounts spanning over 6 months. How do these roles change over time? Do  \textit{sympathizers} get more involved in extremist movements and eventually become \textit{solicitors} or \textit{educators}? To understand, we investigate the role dynamics and ask:   

% Contrary to the popular image of dedicated social movement activists such as Rosa Parks and Cesar Chavez, a more realistic picture of social movement participation is akin to sporadic bursts of inconsistent activity \cite{corrigall2011patterns}. Consequently, next we ask:

\vspace{3pt}
\noindent \textbf{RQ2a:  How long do the extremist accounts retain their original roles?}

\noindent \textbf{RQ2b:  How likely are the extremist accounts in one role to transition to other roles?}
\vspace{3pt}

%Specifically, we measure how long the accounts retain their originally identified roles and the probability by which they might transition into other roles. 
Upon measuring role retention, we find that 66\% of the \textit{solicitors} and 70\% of the \textit{educators} retain their roles throughout our analysis window. When analyzing role transition, we find that \textit{motivators} and \textit{flamers} transition into \textit{sympathizers} with high probability.

As illustrated in our opening examples of \textit{Pissed off White Americans} and \textit{White Lives Matter Movement}, information can be valuable in putting out call for action or educating the readers about extremist ideologies. Researchers argue that information, especially disinformation spreads by flowing strategically or organically through various accounts on social media \cite{starbird2019disinformation}. Specifically, Facebook accounts involved in extremist activities were found to use fake news, biased information and conspiracy sources to promote fundamentalist views \cite{Phadkecross}.
%Another important aspect of online participatory activism is information mobilization through an array of supporters and sympathizers in order to increase the reach of polarizing information \cite{krona20195}.
%In fact, according to the information warfare doctrine released by ISIS in 2016, everyone that participates in the production and delivery of propaganda is regarded as one of the Islamic
%State's ``media mujahidin'' \cite{krona20195,winter2017media}. 
% Other qualitative studies also reveal how information, specifically disinformation, spreads by flowing strategically or organically through various accounts on social media \cite{starbird2019disinformation}. 
Hence, we finally ask:

\vspace{3pt}
%\noindent \textbf{RQ3:  How influential are the roles played by extremist accounts in spreading information from various sources?}
\noindent \textbf{RQ3:  How influential are the roles in spreading various types of information?}
\vspace{3pt}

% Given the prevalence of mis and disinformation in extremist communities \cite{Phadkecross,starbird2019disinformation} we specifically model the information flow for four types of information sources referred from OpenSources.co\footnote{\url{https://github.com/BigMcLargeHuge/opensources}}---hate content, biased news, fake news and conspiracy content. 
To answer, we first use Hawkes process to model the temporal and statistical characteristics of the information spread through the extremist accounts. Next, we use the parameters estimated from the Hawkes process to measure the influence of roles across four types of information sources: extremist, biased, fake news and conspiratorial sources.  We consider a particular role to be \textit{influential} in the spread of information, when a link posted by that role induces an account in another role to post the same link with high probability.
%We consider the temporal flow of a link through various roles and fit Hawkes process for every URL. We investigate which roles are important or influential in the information flow of various types of sources. We consider a particular role to be ``important'' in information spreading when the URL posted by accounts assuming that role induces URL posting by accounts assuming other roles with high probability. 
We find that for information originating from extremist sources, \textit{educators} and \textit{solicitors} are the most influential in triggering other roles to also spread such content. Whereas, \textit{motivators} influence other roles in spreading biased news, \textit{flamers} are influential in the spread of fake news.   

% Overall, our work makes following contributions:
% \begin{itemize}
%     \item We identify five functional roles in online extremist content mobilization based on theories in social movement participation
%     \item We model the hate, fake news, biased news and conspiracy information flow through the roles and identify influential roles in each type of source
% \end{itemize}

Overall our work makes the following contributions:
\begin{itemize}
    \item We offer a framework for systematic operationalization of theoretically motivated characteristics of social movement participation.
    \item We present a data-driven and expert validated taxonomy of social roles in online extremist movements. 
    \item Through a rigorous temporal statistical modeling of information flow, we reveal how different roles are influential in stimulating the spread of extremist, biased, fake and conspiratorial content on Facebook. 
\end{itemize}

Our results allow us to understand how theories of social movement participation are reflected in online extremist movements and where the various roles are located on the trajectories of deeper engagement into extremism. 
Further, our work has implications in assessing which roles might benefit, and which might not, from interventions targeted to counter online extremism. Finally, we discuss how participatory activism might be democratizing the extremist movement and empowering the supporters with the resources and affordances to spread extremism online. 

% The rest of the paper is organized as follows: First, we present the background work situating our research in the related literature. Next, after describing the data, we subsequently discuss methods and results for each of the research questions. We conclude by discussing theoretical and practical implications of our results before presenting the limitations of our work. 

\section{\textbf{Background}}
\subsection{\textbf{Participatory Activism and Extremist Movements}}
Activism means taking an action to effect social change \cite{now2018introduction}. Participatory activism is a kind of activism that is grounded in communities organizing to increase popular support for sociopolitical issues by strategically engaging in both, online and offline mechanisms for participation \cite{rodan2017activism}. Considering the expanse of social media, researchers have connected participatory activism to the concept of ``smart mobs''---people who are able to act in concert even if they don’t know each other \cite{rheingold2007smart,rodan2017activism}. 
Under this new age activism, diverse set of people with overlapping interests can come together in solidarity and act without having to acknowledge who they are beyond what they support \cite{rodan2017activism}. 

How does online participatory activism advance social causes? Previous research presents opposing perspectives on the effectiveness and the legitimacy of the online participatory activism. 
Researchers specifically use the terms ``clicktivism'' or ``slacktivism'' that refer to social media engagement in political micro-action \mbox{\cite{vromen2017digital}} through low cost activities such as liking or sharing the content in order to raise awareness \mbox{\cite{rotman2011slacktivism}}. Writers on popular press comment that ``clicktivism'' or ``slacktivism'' are largely unproductive and ephemeral---an ideal type of activism for lazy generation \mbox{\cite{gladwell2010small,morozov2009brave,white2010clicktivism}}. However, communication and political science scholars argue that such low cost low risk participation is not only widespread but is also becoming a legitimate channel for political activism \mbox{\cite{halupka2018legitimisation}}. Specifically,  Obar et. al. interviewed advocacy groups and found that all the groups viewed online participation as an effective tool for civic engagement and collective action \mbox{\cite{obar2012advocacy}}. Researchers have mostly investigated participatory activism in the context of positive social change. For example, in feminist movements, participatory activism resulted in enhanced understanding of feminism \cite{keller2012virtual} and facilitated legal and public policy discourse surrounding gender based violence \cite{nunez2017online}. In other examples, participatory activism increased the visibility of police brutality against a group of environmentalists \cite{arda2015construction} and enabled successful spread of information in an authoritarian regime \cite{rahimi2016vahid}. While these studies demonstrate how participatory activism can empower populations in social justice causes on one hand, on the other, anti-social movements, for example those advocating for terrorism and extremism, can also benefit from similar practices.  One qualitative study emphasized the success of hashtag campaigns and information manifestos in ISIS's territorial expansion in 2014 \cite{krona20195}. They argued that through numerous online accounts, or ``media operatives,'' ISIS was simultaneously able to promote its positive self-image for recruiting combatants and elicit participation from distant supporters. In other words, the ``media operatives'' played crucial roles in spreading diverse sets of narratives to enable recruitment into ISIS. Moreover, they did so by being a part of strategic information campaigns \cite{krona20195}. For example, ISIS media operatives urged their followers to download videos containing ISIS propaganda and re-upload them on various platforms so as to increase information dissemination while also evading content moderation \cite{krona20195}. 

Based on these studies, it is clear that similar networking and information sharing affordances contribute to both, positive social changes and anti-social movements. 
Yet, studies investigating the darker side of participatory activism are rare. In this paper, we fill this gap by asking: What are the different roles played by extremist accounts in extremist social movements? How stable or transitory are these roles? And how influential are these roles in spreading mis- and disinformation? We specifically focus on U.S domestic extremism, such as white supremacy and anti-LGBT movements and identify various roles through the lens of social movement and resource mobilization theories.  

\subsection{\textbf{Extremism and Social Movements}}
Social movements are collective efforts to bring out the collective action fitting a specific goal or ideology \cite{mccarthy2003social}. Social movements emerge when constitution and function of the society is misaligned with the movement's goals \cite{mccarthy2003social}. By this logic, extremist movements, such as white supremacy,  become more active and aggressive when there is increased racial diversity in the society \cite{zald1966social}. To understand the factors that may lead to the success of extremist movements, we look at theories analyzing the success of social movements. What makes social movements successful?

\subsubsection{\textbf{Success of Social Movements: }}
Researchers attribute the success of social movements to the availability of human, material, and monetary resources \cite{MccarthyResourceTheory}. Human resources include labor, experience, skills, and expertise of the members of the social movement. Material resources include property, office spaces supplies and monetary resources include funds contributed by the members. In other words, by participating in the movement, individuals make their human, material and monetary resources available to the movement \cite{MccarthyResourceTheory}. These resources are are then directed  towards mobilizing supporters, transforming mass public into movement sympathizers and eventually bringing about the desired social change. However, the efficiency with which the resources are translated into action depends on various actors involved, such as volunteers offering human resources or supporters offering monetary resources through fundraising. Based on how various members are involved in the movement, researchers have proposed theoretical roles in social movements participation. Next, we summarize the theoretically proposed roles in social movements and discuss the challenges in adapting them to the online setting. 

\subsubsection{\textbf{Theoretically Identified Roles in Social Movement Participation: }}
Prior scholarly work has described participants in social movements from various theoretical perspectives,  such as participant's role in an organizational hierarchy (e.g., leaders and followers) \cite{morris2002leadership}, their involvement in resource mobilization (e.g., members who distribute resources versus members who consume resources ) \cite{MccarthyResourceTheory} and whether they benefit from the social movement (e.g. stakeholders in the movement) \cite{MccarthyResourceTheory,wahlstrom2018conscience,edwards2004resources,owenconscience}. In organizational hierarchy, \textit{leaders} are strategic decision-makers who mobilize \textit{followers} \cite{morris2002leadership}. 
Considering participants' involvement in resource mobilization, \textit{constituents} provide resources to mobilize \textit{adherents} \cite{MccarthyResourceTheory} and gain sympathy from \textit{bystanders}  \cite{turner1969public}. Moreover, scholars dichotomize the stakeholders of social movements into \textit{potential beneficiaries}: population that directly benefit from the goals of social movement and \textit{conscience participants}: supporters who may not directly gain from the success of the movement \cite{MccarthyResourceTheory,wahlstrom2018conscience,edwards2004resources,owenconscience}. 
How do these theoretically identified roles describe participation on online social media? 
Can we directly adopt the theoretical taxonomy (e.g., \textit{leaders}, \textit{followers}, \textit{constituents} etc.) to describes roles in online extremist movements? 

We identified two concerns with directly adopting a theoretical taxonomy, that consequently motivate the methodologies of our first research question. First, the roles derived from theories of social movements are based on physical social movement participation, such as protest events \cite{wahlstrom2018conscience,owenconscience}. Compared to online participation, physical participation requires increased commitment by members, such as on-the-ground physical presence at the protest events and significant time investment in the movement's activities \cite{rodan2017activism}. On the contrary, in online activism, individuals can become a part of the movement by simply clicking, re-posting or writing short messages on relevant links \cite{rodan2017activism}. Secondly, theoretically identified roles are harder to disambiguate from each other. Consider for example \textit{leaders} defined as per the organizational hierarchy perspective and \textit{constituents} from the resource mobilization perspective. While \textit{leaders} lead the followers, they can also be \textit{constituents}---the distributors of resources. Similarly, both \textit{adherents} and \textit{bystanders} from resource mobilization perspective can be viewed as \textit{followers}. Considering these two points, instead of directly adopting theoretical taxonomy of roles, we consider the underlying characteristics of participation. Based on the framework of features derived from the theories of social movement participation,  we develop a new taxonomy of roles for the online setting. Next, we detail the underlying characteristics that form a background for our role identification process. 

\subsection{\textbf{RQ1 Background: Characteristics of Social Movement Participation}}
We summarize the theories in social movement participation across three dimensions: \textit{drives for participation}, \textit{engagement in the movement}, \textit{strategies of mobilization}. These three dimensions, and the computational features derived from them, are at the crux of our methodology for identifying roles in the extremist movements. The first two columns in Table \ref{tab:feature_table} summarize the models described in the next subsections.

\subsubsection{\textbf{Drives for Social Movement Participation:}}
\label{sec:drives_lit}
%Different types of actors can have different drives or motivations to participate in online movements. 
Theories of drives behind social movement participation can be broadly distilled in two groups---expectancy-value models and social-psychological models \cite{ferree1985mobilization}. The expectancy-value models stress on the rationality of the participants whereby the individuals weigh costs and benefits of participation while decision making \cite{klandermans1984mobilization}. Particularly, they consider the risk-reward ratio of the involvement in the social movement before investing and mobilizing resources for their own benefit \cite{oberschall1973social,marx1975strands}. Scholars pointed out one limitation of expectancy-value model that it underestimates the role of ideological drives and shared grievances in participants \cite{klandermans1984mobilization}. To fill this gap, social-psychological models attribute the movement participation to various psychological drives. Feelings of injustice, relative deprivation and moral outrage is at the heart of movement participation \cite{van2013socialprotest}. While such grievances related to social movement issues are ubiquitous, not all of them turn into protests \cite{van2013social}. Hence, social movement scholars also consider the sense of efficacy or achievement that drives people to participate \cite{van2013social}. For example, people might participate because they believe that their collective action can actually achieve the social change \cite{gamson1992talking}.  Further, researchers also stress the importance of group identity. The more people identify with the social groups involved in the movement, the more they are inclined towards participating in the movement \cite{simon2001toward}. Emotions also play an important role in driving people into the social movements. For example, anger is considered as \emph{the} prototypical emotion for protests \cite{van2017individuals}.

\subsubsection{\textbf{Engagement Trends in Social Movements: }}
\label{sec:engage_lit}
%Social media allows voluntary participation from actors in the social movement. 
Various types of participants could adopt different degrees of engagement and commitment to distributing social movement related resources. 
%hate related content based on their interests. For example, leading organizations in the anti-LGBT movement may use social media to primarily promote anti-LGBT agenda resulting in high engagement and commitment to distributing anti-LGBT information. Whereas, groups that merely support the anti-LGBT stance might engage with hate content less frequently. 
Specifically based on the availability of resources, \textit{constituents} actively distribute the resources in order to proselytize the \textit{adherents} and bystanders  \cite{MccarthyResourceTheory}. Moreover, actors could participate in resource mobilization for a single or multiple social movements \cite{MccarthyResourceTheory}. Another important aspect of the engagement is how it varies over time. People participate in the social movements with varying degrees of continuity. Corrigall-Brown characterizes such periodic engagement across four dimensions---persistence, transfer, abeyance, disengagement \cite{corrigall2011patterns}. This characterization of participation trajectories is especially important on social media because it accrues diverse group of individuals with varying degrees of interests and dedication. 
%Thus, focusing only on the highly engaged committed activists may undermine the broader picture of extremist information mobilization online.

\subsubsection{\textbf{Strategies of Information Mobilization: }} 
\label{sec:strategy_lit}
 Social movements are goal-oriented. Participants in the social movements strategize the distribution and uses of resources to induce collective action and gain support \cite{MccarthyResourceTheory}. Specifically, core members of the social movement mobilize resources in order to recruit volunteers, collect funds, spread their agenda and hold gatherings. Other researchers also find such solicitation strategies to be crucial for financing the social movements \cite{bromley1980financing}. While solicitation may directly affect the progress of the movement, researchers also highlight resources that can indirectly create opportunities for collective action. Specifically in the social media setting, expressing opinions, thoughts and beliefs around political events \cite{valenzuela2013unpacking} and reporting events to increase users’ knowledge of public issues, political causes, and social movements \cite{valenzuela2013unpacking,de2006news} are considered as key strategies of online activism. 
 
 %Understanding how various accounts use the extremist content, e.g. for expressing opinions or soliciting, can also reveal the roles they play in extremist information sharing. 
% \subsection{\textbf{}}

% \subsection{\textbf{}}

%\section{\textbf{Extremism as Social Movement}}
%  In this section we first provide the literature review for extremism on social media and extremist information operations. Next, we consider information sharing as a valuable resource for online extremist movements and also provide background on traditional roles in social movement participation via the resource mobilization perspective.  

\subsection{\textbf{Role of Information in Online Extremist Movements }} 
%Like other social movements, the rise and fall of extremist movements depend on the voluntary participation by their supporters and their organizational structure \cite{freilich2009critical}. 
Extremist movements have started leveraging social media to increase their outreach \cite{johnson2019hidden,Extremis62online}. According to Anti Defamation League (ADL) \cite{adlabout}---an anti-hate organization focusing on combating anti-semitism in USA---2018 was recorded at the most violent year in terms of deaths caused by domestic extremism since 1995 \cite{RightWin18online}. Several popular social media platforms were criticized in these attacks for enabling recruiting and spreading disinformation \cite{formica2020social}. For example, the entire Christchurch mass shooting seemed to be orchestrated for the social media age \cite{HowtheCh44online}. The perpetrator of the mass shootings posted an 87 page manifesto on 8chan with anti-immigrant and anti-Muslim ideas and directed readers to a Facebook page where the shootings were live streamed. 
This is one of the many examples where extremists have used social media to broadcast their ideologies followed by violent events. Apart from educating the public about the extremist agenda, extremists also use online platforms to promote action \cite{muller2020fanning}. Recently, the right-wing extremists have started leveraging meme and trolling culture to shape hate narratives \cite{DataSoci51online}. Researchers found that extremist fringe groups collaboratively launder information even on mainstream social media platforms to appeal to the mass \cite{Phadkecross}; especially to a young audience \cite{davey2017fringe}. 

% \subsection{\textbf{Extremist Information Operations: }}
% Information operations involve the dissemination of information with the intention of achieving a competitive advantage over an opponent \cite{vitali2020fueling}. Previously, researchers have explored information operations by racial justice actors \cite{arif2018acting}, foreign adversarial actors such as Russian bots and trolls \cite{broniatowski2018weaponized,hagen2020rise,zannettou2019disinformation} and political and medical disinformation campaigns \cite{starbird2019disinformation,xaudiera2020ibuprofen,mckay2020disinformation,vargas2020detection}. Starbird et. al. \cite{starbird2019disinformation} reveal the ``participatory'' nature of the online political disinformation campaigns. Meaning, they state that political disinformation can strategically or organically flow through various online communities and actors who might or might not be aware of their role in the information operation. We take inspiration from their findings and also characterize the participatory nature of online extremist movements by finding various emergent roles. 

Despite the strong visibility of extremist groups online, computational studies focusing on information operations of U.S. domestic extremist groups are rare. A recent qualitative work investigated information sharing by various extremist ideologies and find that hate groups use information sharing to recruit, radicalize and educate their follower base \cite{Phadkecross}. Specifically, they observed the use of links shared from biased news sources by white supremacy groups and links from websites hosting conspiratorial content in anti-Muslim  and religious supremacy groups.
In this paper we use rigorous temporal statistical models to quantify how influential various roles in extremist movements are in disseminating different types of information sources such as extremist content, fake news, biased news and conspiracies.

\section{\textbf{Data}}

% \begin{figure*}[t]
%     \centering
%     \includegraphics[width=0.60\textwidth]{figures/page_example.pdf}
%     \caption{Facebook page with highest Page Likes in our dataset. (a) page description claiming to provide reliable commentary on current events and (b) NewsGuard rating for the page indicating that the websites fails to adhere basic journalistic standards. }
%     \label{fig:page_example}
% \end{figure*}

% \usepackage{booktabs}
% \usepackage{graphicx}
\begin{table}[]
\centering

\resizebox{0.45\textwidth}{!}{%
\begin{tabular}{@{}lllll@{}}
\toprule
 \textbf{statistics (per account)}& \textbf{min} & \textbf{max} & \textbf{mean} & \textbf{std.} \\ \midrule
\textbf{posts} & 71 & 932K & 7,067 & 30,574 \\
\textbf{link posts} & 23 & 78,571 & 1,614 & 3,915 \\
\textbf{extremist link posts} & 10 & 5,129 & 207 & 528 \\
\textbf{engagement} & & & & \\
\hspace{4pt} page likes & 106 & 1.8M & 7,241 & 36,576 \\ 
\hspace{4pt} group members & 35 & 2.2M & 2,827 & 13,603 \\\bottomrule
\end{tabular}%
}
\caption{Descriptive statistics for extremist accounts in our dataset. There are a total of 4,876 extremist accounts in our dataset.}
\label{tab:data_description}
\end{table}

In this work, we focus on identifying roles in online extremist movements and assess how influential various roles are in spreading information from extremist, biased, fake news and conspiratorial information sources. Specifically, we focus on \textbf{\textit{extremist accounts}}---public Facebook pages and groups that share links from extremist websites. In this section we detail our process of identifying the extremist accounts (Figure \ref{fig:method} (a) and (b)) and collecting their Facebook activity data. We first explain the problem of extremism on Facebook and then describe the data collection and pre-processing steps. 

\subsection{\textbf{Extremism on Facebook Groups and Pages}}
Despite the policies against extremist content, Facebook is still crowded with groups and pages circulating and discussing violent ideas \cite{Facebook32online}. Very recently, Facebook banned a number of pages and groups involved in the ``boogaloo movement''---an anti-government movement by right wing extremists organizing for armed revolt \cite{bogaloo}. However, this ban was not as effective as initially perceived. The pages and groups related to the boogaloo movement simply renamed and rebranded themselves  with harmless-sounding names and continued their extremist activities \cite{techtransparency}. According to the  International Association of Chiefs of Police---a non-profit organization that is the world's largest professional association for police leaders---Facebook groups and pages are at the center of extremist recruitment, radicalization, and mobilization \cite{IACP}. They found that, by just re-posting and linking information from websites hosting 
graphic videos and other violent content, extremism related groups and pages are able to abide by Facebook's policies against hate speech and still spread relevant information. 
%Moreover, the recent moderation efforts by Facebook focused on the ``elites'' of the extremist community, still allow information from their websites to spread across Facebook groups and pages. 
% For example, Virginia Dare, a Southern Poverty Law Center (SPLC) designated white nationalist group \footnote{\url{https://www.splcenter.org/fighting-hate/extremist-files/group/vdare}}, was banned from Facebook in May 2020. However, several Facebook pages and groups share articles from websites that host content from Virginia Dare (VDARE.com) (see Figure \ref{fig:vdare_example}). In other words, even after banning prominent extremist organizations (such as VDARE) that champion extremist causes, Facebook still has an array of groups and pages that launder the information originally published by banned accounts. 
In our first research question, we identify the roles played by various such extremist accounts in advancing the extremist movements online. In order to model these roles, we first need to identify the extremist websites, and then select the extremist accounts---Facebook pages/groups that share links from the extremist websites.  

\subsection{\textbf{Identifying Extremist Websites}}
\label{sec:identify_extreme_domains}
To identify extremist websites,  we take help of the resources published by external organizations who are experts in social justice causes. 
Specifically, we refer to Southern Poverty Law Center's (SPLC) website \footnote{\url{https://www.splcenter.org/}}. SPLC is a non-profit organization engaged in legal advocacy for social justice issues. Every year, SPLC releases a extremist groups' dataset \footnote{see ``DOWNLOAD DATA'' in \url{https://www.splcenter.org/hate-map}} that records the names, locations, and ideologies of extremist groups operating in the United States. We searched through SPLC's 2018 and 2019 extremist groups datasets. 
For each listed extremist group, we manually searched for their official website. Note that each listed extremist group has physical headquarters across various states in the USA. By searching for their websites we identify the home for the extremist groups' content in the online world.  We identified 289 websites hosted by the listed extremist groups. For each website, we also note the website domain---hereafter referred to, as extremist domains. For example, for Virginia Dare, a white supremacy group, we record {\small \tt vdare.com} as a website domain and for Alliance Defending Freedom---anti-LGBTQ advocacy organization---we record {\small \tt adflegal.org}. 
%This dataset includes groups adhering to White Supremacy, anti-Immigration, anti-LGBT, anti-Muslim and religious supremacy ideologies.
%We use a list of hate groups published by Southern Poverty Law Center (SPLC) to identify websites containing extremist content. 

According to the SPLC's policies, SPLC prioritizes identifying all U.S based hate groups regardless of the group's ``left'' or ``right'' political leaning \footnote{\url{https://www.splcenter.org/20200318/frequently-asked-questions-about-hate-groups}}.
%Following SPLC's policies, we also do not focus on extremism related to any particular political ideology. 
For example, the 2019 SPLC dataset contains 27 groups with Black Separatist ideology which is not on the far-right of the U.S political spectrum. However, we could identify websites for only 10 out of 27 Black Separatist groups. While we did not set out to study online extremist groups only from the far-right political spectrum, our collection of extremist websites largely belong to far-right groups representing anti-Immigration, anti-Muslim, White Supremacy and anti-LGBTQ ideologies.  This skew towards far-right extremism is representative of the picture of domestic extremism in the United States. According to the Center for Strategic and International Studies (CSIS), \footnote{CSIS conducts policy studies and strategic analyses of political, economic and security issues throughout the world. CSIS is labeled as least biased and highly factual source of information by mediabiasfactcheck.com} far-right extremism has massively outpaced far-left and other types of extremism in the United States \mbox{\cite{jones2020escalating}}. Another independent report on Global Terrorism Index produced by the Institute for Economics and Peace reports that the far-right attacks in  In North America, Western Europe, and Oceania have increased by 250\% since 2014, making it more lethal than the far-left extremism \mbox{\cite{GTI2020w13online}}. While we believe that our dataset is reflective of the ecosystem of domestic extremism in the United States, we discuss this skew further with respect to extremism in other countries in the Limitations and Future Directions section.

%Next, we want to identify Facebook pages and groups share extremist content.  
% In this work, we focus on identifying roles and modeling information flow through these roles in online extremist movements. We specifically investigate U.S domestic extremism and extremist ideologies such as white supremacy, anti-LGBTQ, anti-Immigration etc. We analyze the activity of various Facebook groups and pages involved in sharing extremism related information. 
%In this work, we focus on understanding the information mobilization in online extremist communities, specifically, Facebook.

\subsection{\textbf{Identifying Extremist Accounts}}
To identify extremist accounts---Facebook groups/pages that share links from extremist websites---we use the CrowdTangle Link Search API \footnote{\url{https://github.com/CrowdTangle/API/wiki}}. The CrowdTangle Link Search API retrieves posts by public Facebook groups and pages containing a certain link or the link domain. With the 289 identified extremist domains, we query the Link Search API separately for every domain. For extremist websites containing generic domains such as \mbox{\small \tt sites.google.com}, we query the full link with the sub-domain, for example, \mbox{\small \tt sites.google.com/site/newblackliberationinstitute}. For every queried domain, the API returns up to 1000 posts containing that link domain. Hence, to increase the completeness of our data, we queried every extremist domain separately for every calendar month starting from January 2018 to December 2019. For every queried domain in any calendar month, the number of returned posts was always less than 1000. This indicates that we have retrieved all public posts on Facebook available to CrowdTangle between 2018 and 2019 that share links from the identified extremist domains. 
%Next, we used the CrowdTangle Link Search API to identify Facebook pages that share links belonging to the hate group websites. 
%The Link Search API takes a link domain as an input and returns a maximum of 1000 posts containing the queried domain. In order to increases the completeness of our data collection, we queried for every domain separately in every calendar month starting from January 2018 to December 2019. 
Every post retrieved from CrowdTangle contains the account (page or group) name, post text, embedded links, timestamp, reactions (e.g., Like, Love, HaHa etc.), number of comments, number of public shares and 12 other fields. 
We aggregate all returned posts (450K posts) and find that our data contains 71,430 unique Facebook pages/groups accounts. In other words, 71,430 accounts shared at least one link from the extremist domain. User activities on social media often follow skewed, long-tailed distributions where most users contribute less frequently while fewer users are more active. We observe similar distribution with extremist links posted per account. Previously, researchers have used activity thresholds to eliminate accounts or communities that are less active \mbox{\cite{Hamilton2017LoyaltyCommunities,kumar2019predicting}}. For example, while studying Wikipedia edits, Kumar et. al. remove the users that make less than 5 edits  \mbox{\cite{kumar2019predicting}}. We decide the activity threshold by analyzing the percentile values of the extremist links per account distribution. 
Based on the $95^{th}$ percentile cut-off, we remove all accounts that share less than 10 unique links from extremist domains. We provide additional details of the downstream analysis with various link thresholds in the Appendix.
The entire data collection process happened across three weeks in May 2020. This means that the extremist accounts that were active in 2018 and 2019 but got banned before May 2020 are not included in the dataset. \footnote{As of January 2021, 207 extremist accounts from our dataset (156 Facebook groups and 51 Facebook pages) have been removed from Facebook.}
%To eliminate accounts that only rarely post links from extremist websites, we removed all accounts that share less than 10 unique links from the extremist domains. 
Finally, we have 4,876 Facebook pages/groups remaining in our dataset. Table \ref{tab:data_description} displays the descriptive statistics for the accounts in our dataset. 

\subsection{\textbf{Qualitative Validation of the Extremist Accounts }}
While we know that the extremist accounts posted 10 or more unique links from the extremist websites, do they promote extremist worldviews in general? In this subsection, we present our qualitative validation of the ideologies and the views promoted by the extremist accounts. To validate, we invited two experts from the Southern Poverty Law Center specializing in white supremacy and anti-LGBTQ hate groups. We requested the experts to qualitatively analyze a random sample of extremist accounts. Specifically, we randomly sampled 20 extremist accounts that share links from the white supremacy and 20 accounts that share links from anti-LGBTQ extremist websites. Next, we asked the experts to review Facebook timelines of the extremist accounts and describe their ideologies based on the content hosted and the page/group name and description. 16 of the 20 accounts posting links from white supremacy extremist websites, generally promote either far-right conspiratorial views and racists or misinformative content. Similarly, 18 of the 20 accounts posting links from anti-LGBTQ extremist websites usually peddle anti-choice, anti same sex marriage or anti-trans views. Only 2 of the 40 groups focused exclusively on memes without actively promoting any aspect of the white nationalist or anti-LGBTQ rhetoric. 
The complete expert analysis is available at the link in the footnote \footnote{\url{https://www.dropbox.com/s/mf3jt9xbk9z9xfw/SPLC_annotations.pdf?dl=0}}. 
%Overall, the expert commented that the extremist accounts hosted content with varying degrees of connection to the white supremacy and anti-LGBTQ ideologies. 
%Many accounts promoted misinformation and far-right conspiracies and a few hosted racist, pro-segregation, anti-trans and anti-abortion content. 

While the experts validated the extremist views and the content hosted by extremist accounts, the 4,876 extremist accounts in our dataset consist of Facebook pages and groups that engage multiple Facebook users. We treat every page or a group as one extremist account that hosts content posted by it's page owners or the group members. Who contributes to the content on these pages or groups? CrowdTangle, or any other Facebook API does not disclose personally identifiable information even in the public posts \footnote{\url{https://help.crowdtangle.com/en/articles/1140930-what-data-is-crowdtangle-tracking}} \footnote{\url{https://developers.facebook.com/docs/groups-api/}}. In other words, when requesting the content's of a post, the response will not include the name of the member who created the post. This poses a challenge in understanding the agency of posts shared on the extremist accounts. While this is a limitation of the Facebook dataset, we approximate the engagement with extremist accounts by reporting the distribution of page likes and group members in Table \mbox{\ref{tab:data_description}}.

%Next, we invited two experts from the Southern Poverty Law Center specializing in white supremacy and anti-LGBT hate groups to analyze a random sample of extremist accounts for relevancy. Specifically, we randomly sampled 40 extremist accounts that predominantly share links from the white supremacy and anti-LGBT hate groups and asked the experts to write the descriptions for the accounts commenting on the content shared and its relevancy to the two extremist ideologies (see the complete expert analysis here \footnote{\url{https://www.dropbox.com/s/mf3jt9xbk9z9xfw/SPLC_annotations.pdf?dl=0}}). This feedback from the expert not only determined the relevancy of the accounts to extremist ideologies, but also motivated our research question identifying the roles played by extremist accounts.} 

% Figure \ref{fig:page_example} displays the information panel of the Facebook page with highest likes. This page belongs to a news website that is deemed unreliable by NewsGuard \footnote{\url{https://www.newsguardtech.com/}}

\begin{figure*}[t]
    \centering
    \includegraphics[width=0.99\textwidth]{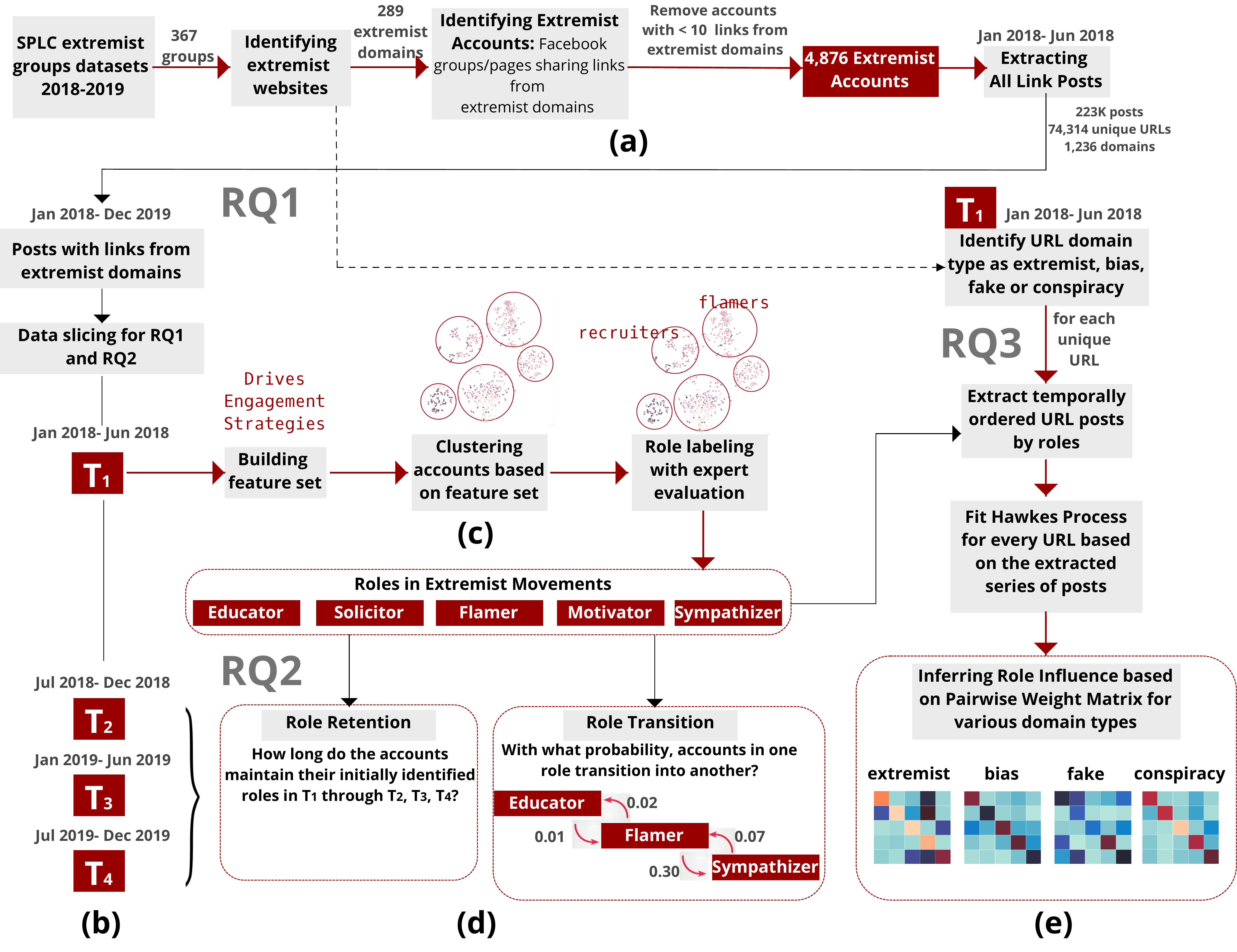}
    \caption{Figure showing data preparation and method details for three research questions. (a) We first identify extremist websites (domains) and identify extremist accounts---Facebook pages and groups that links from extremist domains. Next, we extract posts by the extremist accounts that contain links from the extremist websites and (b) slice the dataset into four time windows of 6 months each. (c) For role identification in RQ1 we cluster extremist accounts based on the features derived from the account activity in T1. (d) We use the cluster centers from T1 to re-cluster the extremist accounts and measure role retention and transition in RQ2 (also see Figure \ref{fig:rq2_method} for more details). (e) For RQ3, we use the link posts by the extremist accounts made in $T_{1}$, group the links based on their source type (extremist, biased, fake news and conspiracy) and measure the influence of various roles in link sharing based on Hawkes process.}
    \label{fig:method}
\end{figure*}

\subsection{\textbf{Extracting Information Shared by Extremist Accounts}}
\label{sec:sourcetype}
We identified 4,876 extremist accounts. In our third research question, we assess how influential various accounts are in spreading links from biased, fake news and conspiracy domains in addition to the extremist domains. Towards this goal, we need all \textit{link posts}---posts with embedded links---from the extremist accounts and not just the ones originating from extremist domains.  Hence, next we extract all link posts made by the extremist accounts between 2018 and 2019 and then group the link posts by the link domain type.   
First, we use the CrowdTangle Post Search API to acquire all of the Facebook link posts made by the 4,876 extremist accounts. 
%\hl{We observed that a large amount of link posts contained links to popular social media platforms such as \mbox{\tt \small facebook.com}, \mbox{\tt \small twitter.com}, \mbox{\tt \small reddit.com}, \mbox{\tt \small youtube.com} etc. Since such links belong to generic social media domains, it is not possible to approximate the type of content they might host. Hence, we first manually identified 20 most frequent social media domains in our dataset and removed links belonging to those domains.}
In total, we obtain 223K link posts made by extremist accounts across two years, 2018 and 2019. Next, we extract the links shared in the posts along with the timestamps and the link domains. In total there are 74,314 unique links in our dataset spanning over 1,236 link domains. 

%How can we identify whether the link domains host extremist, biased, fakes news or conspiracy content? 

% In RQ3, we use the timestamped link data and the domains to assess the importance of the accounts in sharing extremist content, biased news, fake news and conspiracy sources. 

\subsection{\textbf{Data Slicing for the Downstream Analysis: }}
Our role identification process discussed in next sections is based on the Facebook activity of extremist accounts. Recall that we have two years (2018 and 2019) of activity for each account. Here, we determine the time unit of analysis in which we identify the roles, observe their dynamics and assess their influence in information sharing.
%Using all of this activity to identify the role played by an account in extremist information mobilization will prevent us from studying if the account transitions into other roles over time. 
To study the role stability over time, we need to divide the two year timespan into smaller windows and analyze how extremist accounts transition into different roles over successive time windows. We slice the data into four windows of six months---$T_{1}$, $T_{2}$, $T_{3}$, $T_{4}$--from January 2018 to December 2019 (Figure \ref{fig:method} (b)). In RQ1, we identify the roles played by extremist accounts in $T_{1}$ and in RQ2, we track the role stability over $T_{2}$, $T_{3}$ and $T_{4}$. In RQ3, we model the information flow through the roles in $T_{1}$. Does the choice of six-month time window affect our observations? 
We experimented with different length of time windows (specifically, 2 months, 3 months and 12 months) obtaining similar results for the research questions.

%\vspace{3pt}
\subsection{\textbf{Identifying Domain Types for Extracted Links:}}
In the RQ3 analysis, we model the information flow and examine the influence of various roles in spreading content from mis and disinformation based sources. Specifically, we consider 57K link posts made in $T_{1}$ and identify domain types for the links.
%With 223K link posts extracted containing 74,314 unique links spanning over 1,236 link domains, we now need to identify domain types. 
We are interested in identifying domains hosting extremist content, biased news, fake news and conspiracy content. We refer to the dataset published by OpenSources \footnote{\url{https://github.com/BigMcLargeHuge/opensources}}. OpenSources maintains a professionally curated list of online sources towards the goal of empowering people with reliable information. With the help of professionals, they annotate each source based on overall inaccuracy, extreme biases, lack of transparency, and writing styles. We use the annotations from OpenSources, and our list of originally identified extremist domains (Section \ref{sec:identify_extreme_domains}), to  group links by their source type as extremist, fake news source, biased source and conspiracy/pseudoscience source. For our RQ3 analysis, we remove all links that do not belong to either of these four domain types. Removing the links with domains not included in the OpenSources or the original set of extremist domains, reduces our $T_{1}$ link post dataset by 24\%. In the Appendix, we provide the characterization of the frequent domains in the removed dataset.

\section{\textbf{RQ1 Method: Identifying Roles in Online Extremist Movements}}
In this research question, we identify roles played by extremist accounts in the extremist movements based on theory guided characteristics of social movement participation. Specifically, we operationalize \textit{drives for participation} (Section \mbox{\ref{sec:drives_lit}}), \textit{engagement in the movement} (Section \mbox{\ref{sec:engage_lit}}) and \textit{strategies of mobilization} (Section \mbox{\ref{sec:strategy_lit}}) based on the activity of extremist accounts in the six month time period $T_{1}$ (Jan 2018-Jun 2018) and build a feature set (Table \mbox{\ref{tab:feature_table}}) to identify roles. Next, we represent every extremist account in our dataset with the derived features. To identify roles, we cluster the accounts based on the derived features. Finally, we qualitatively analyze and label each cluster as a role in extremist social movement with the help of experts in social psychology. 

\subsection{\textbf{Operationalizing Characteristics of Social Movement Participation}}

\begin{table*}[]
\centering
%\scriptsize 
\sffamily 
\resizebox{\textwidth}{!}{%
\begin{tabular}{@{}lllll@{}}
\toprule
\textbf{\begin{tabular}[c]{@{}l@{}}Characteristics of\\ Participation\end{tabular}} & \multicolumn{1}{l}{\textbf{Theoretical Models}} & \textbf{References} & \textbf{Behavior} & \textbf{Operationalization} \\ \midrule
\multirow{6}{*}{\begin{tabular}[c]{@{}l@{}}\textbf{Drives for}\\ \textbf{participation}\end{tabular}} & \multicolumn{1}{l}{\multirow{2}{*}{\begin{tabular}[c]{@{}l@{}}Expectancy-value\\ models\end{tabular}}} & \cite{klandermans1984mobilization,oberschall1973social,marx1975strands} & Risk & Proportion of LIWC  Risk words ({\small e.g., caution, crisis, failure}) \\ \cmidrule(l){4-5} 
 & \multicolumn{1}{l}{} & \cite{klandermans1984mobilization,oberschall1973social,marx1975strands} & Reward & Proportion of LIWC  Reward words ({\small e.g., benefit, bonus, award}) \\ \cmidrule(l){2-5} 
 & \multirow{4}{*}{\begin{tabular}[c]{@{}l@{}}Social-psychology \\ models\end{tabular}} & \cite{van2013socialprotest} & Injustice & Proportion of MFD Fairness words ({\small e.g., parity, fair, justice})\\ \cmidrule(l){4-5} 
 &  & \cite{van2013social,gamson1992talking}  & Achievement & Proportion of LIWC Achievement words ({\small e.g., accomplish, ability, attain})\\ \cmidrule(l){4-5} 
 &  &\cite{simon2001toward}  & Group Identity & Proportion of LIWC  we words ({\small e.g., we, ours, us})\\ \cmidrule(l){4-5} 
 &  &\cite{van2017individuals}  & Anger & Proportion of LIWC anger words ({\small e.g., resent, argue, angry}) \\ \midrule
\multirow{5}{*}{\begin{tabular}[c]{@{}l@{}}\textbf{Engagement in} \\ \textbf{the movement}\end{tabular}} & \begin{tabular}[c]{@{}l@{}}Degrees of \\ participation  \end{tabular} & \cite{MccarthyResourceTheory} & \begin{tabular}[c]{@{}l@{}}Proportion of links \\ from extremist domains\end{tabular} & \begin{tabular}[c]{@{}l@{}}Ratio of links from extremist domains \\ to total link posts\end{tabular} \\ \cmidrule(l){2-5} 
 & \multirow{3}{*}{\begin{tabular}[c]{@{}l@{}}Degrees of\\ participation (popularity)\end{tabular}} & \cite{MccarthyResourceTheory} & Likes & \begin{tabular}[c]{@{}l@{}}Proportion of likes on extremist links \\ to likes on the rest of the link posts\end{tabular} \\
 &  &  & Shares & \begin{tabular}[c]{@{}l@{}}Proportion of shares on extremist links \\ to likes on the shares of the link posts\end{tabular} \\
 &  &  & Comments & \begin{tabular}[c]{@{}l@{}}Proportion of comments on extremist links \\ to comments on the rest of the link posts\end{tabular} \\ \cmidrule(l){2-5} 
 & \begin{tabular}[c]{@{}l@{}}Trends in  \\ participation \end{tabular} & \cite{corrigall2011patterns} & Trend & \begin{tabular}[c]{@{}l@{}}Trend line fitted on the number of extremist \\ links posts per month\end{tabular} \\ \midrule
\multirow{2}{*}{\begin{tabular}[c]{@{}l@{}}\textbf{Strategies of} \\ \textbf{ mobilization}\end{tabular}} & Opinions & \cite{valenzuela2013unpacking} & Expressions of opinions & \begin{tabular}[c]{@{}l@{}}Proportion of extremist link posts \\ containing opinion patterns (see Table \ref{tab:op_pat} )\end{tabular} \\ \cmidrule(l){2-5} 
 & Solicitation &\cite{MccarthyResourceTheory,bromley1980financing}   & Expressions of solicitation & \begin{tabular}[c]{@{}l@{}}Proportion of extremist link posts \\ containing solicitation patterns (see Table \ref{tab:sol_pat} )\end{tabular} \\ \bottomrule
\end{tabular}%
}
\caption{Table summarizing features used to identify roles in online extremist movements on Facebook. We build the feature set based on underlying characteristics of participation and the theoretical models describing them. }
\label{tab:feature_table}
\end{table*}

\subsubsection{\textbf{Drives for Social Movement Participation (6 features).}}
Theoretical models exploring the drives for social movement participation consider two distinct perspectives: \emph{expectancy-value models}, whereby participation is driven by perceived risk-reward assessments and \emph{social psychology models}, where the psychological features are the core drivers of participation. Below we detail the computational operationalizations of \emph{drives} informed by these theoretical models explained in Section \ref{sec:drives_lit}.
%We provide the details of the theoretical models describing drives for social movement participation in Section \ref{sec:drives_lit}. 
%Here we briefly restate the key points and detail their computational operationalizations. 
\begin{itemize}
    \item \textbf{Expectancy-value features: } Participation in social movement could be driven by perceived risk-reward assessment related to engagement in the movement \cite{klandermans1984mobilization,oberschall1973social,marx1975strands}. Analyzing the language used by the extremist accounts while sharing links from extremist websites, especially the words related to cost-benefit, could indicate whether the accounts considered the costs and benefits of participating in extremist movements. To operationalize these features, we use risk and reward lexicons from the Linguistic Inquiry and Word Count (LIWC) 2015 \cite{tausczik2010psychological}. LIWC is designed to record words that reflect various psychological states and perceptions \cite{tausczik2010psychological}. 
    Specifically, we calculate the proportion of risk related words (e.g., caution, crisis, failure) and reward related words (e.g., benefit, bonus, award) used by an extremist account while sharing links from extremist websites. 
    \item \textbf{Social Psychology features: } Guided by the social-psychology based theoretical models, here we want to measure the feelings of injustice \cite{van2013socialprotest} , sense of achievement \cite{van2013social,gamson1992talking}, group identity \cite{simon2001toward} and anger \cite{van2017individuals} that could serve as potential drives for participation. To identify the language related to injustice, we use Moral Foundations Dictionary that contains a systematically derived list of words pertaining to moral foundations in political ideologies \cite{takikawa2019moral}. Specifically, we use the ``fairness'' lexicon which accommodates virtue words such as rights and equality, and vice related words such as bigot, favoritism, and prejudice \cite{frimer2019moral}. Next, to measure the sense of achievement, we use LIWC's \cite{tausczik2010psychological} achievement category which contains words such as ``accomplish'', ``ability'', ``attain'' etc. Further, language related to group identity can be reflected by the use of third person pronouns such as ``we'', ``us'', ``ours'' \cite{job2004shared,tausczik2010psychological,pavalanathan2015identity}. Hence we measure the third-person pronoun usage by LIWC ``we'' category. Finally, to measure anger related words, we use LIWC anger category containing words such as ``resent'', ``argue'', ``angry.'' For each of these lexicons, we calculate the proportion of words in each lexicon while sharing links from extremist websites.  
\end{itemize}

\subsubsection{\textbf{Engagement Trends in Social Movements: (5 features). }}
As per the details described in Section \ref{sec:engage_lit}, here we provide our methods to characterize engagement trends in the social movements. Participants can engage with social movements in various degrees of interests and continuity \cite{MccarthyResourceTheory,corrigall2011patterns}. Hence, we calculate proportion, popularity and trends in sharing links from extremist websites. 

\begin{itemize}
    \item \textbf{Proportion of Links from Extremist Domains: } Various degrees of participation in distributing resources can reflect the involvement in social movement \cite{MccarthyResourceTheory}. Hence, for every extremist account, we first calculate \textit{proportion of links from extremist domains}---proportion of links shared from extremist domains to all links shared by that account in a given time-frame.
    \item \textbf{Popularity of Links from Extremist Domains: } %The Facebook pages and groups in our dataset represent the thoughts of collective of people. 
    The amount of positive reactions and interactions on the shared links could reflect how popular the posts containing extremist links are on a Facebook page/group.  Hence, we calculate the average likes, shares and comments received on posts with links from extremist websites and divide it by the average likes, shares, and comments (respectively) on all links posted in a given time-frame. High values of these features indicate that the extremist content is more popular compared to the rest of the content published on that page/group.
    \item \textbf{Trends in Disseminating Links from Extremist Domains: } We also account for the engagement trends. For each month within the six-month period, we calculate the number of links from extremist websites posted on an extremist account and fit a line via least-square regression. Least-square regression finds optimal fit for the line by minimizing the sum of squared residuals.
    We calculate the trend of this fitted line to measure engagement. Specifically, positive values of trend can indicate increasing engagement and negative values can indicate disengagement in posting links from extremist websites. 
\end{itemize}

\subsubsection{\textbf{Strategies of Information Mobilization: (2 features): }}
As described in detail in Section \ref{sec:strategy_lit}, core members of the social movements may strategically solicit participation through calls for donations, volunteers and invitations for social gatherings \cite{MccarthyResourceTheory,bromley1980financing}. Similarly, members can also strategically create opportunities for collective action by expressing opinions, thoughts and beliefs around political events \cite{valenzuela2013unpacking}. Hence, we build two features that capture the expressions of personal opinions and the language of solicitation used by extremist accounts while sharing links from extremist domains. For both features, we calculate the proportion of extremist link posts containing the expressions of opinions and solicitations respectively. 

%We observed from our data that users often employed specific language strategies while posting the links from extremist domains. Specifically, we looked at the random sample of 500 link posts to understand various communication strategies. Most of the links contained reporting of events or news. Some used the information from links to voice their opinions about political and social issues. Some used the posts to solicit participation from other users. To capture these nuanced intentions, we built various phrase patterns that could signal opinion expression and solicitation. Specifically, through iterative annotations we build a high precision low recall system for detecting opinions and solicitation pattern in the hate link sharing. 

\begin{itemize}
    \item \textbf{Expressions of Opinions in Posts with Links from Extremist Websites: } By ``opinions'' we refer to the expression of thoughts, beliefs and personal opinions \cite{sauri2008factuality}. To calculate the proportion of opinions present in posts, we extract phrases that signal expressions of personal opinions or private states \cite{wiebe2005annotating}. 
    Previously, researchers studying emotional and informational support \cite{biyani2014identifying,wang2012stay} relied on emotional and informational support related nouns, verbs and  adjectives to extract phrases related to their task. For example, Wang et. al. \cite{wang2012stay} used $<you + MODAL VERB>$ pattern to extract phrases containing suggestions (``you should'' or ``you must''). We use similar methodology to extract phrases related to personal opinions. How can we identify verbs, nouns and adjectives related to personal opinions? Chen et. al. argue that individuals form their opinions via cognition and internal perceptual cues \cite{chen2019opinion}. Hence, we first look at LIWC 2015 cognitive processing lexicon and its subcategories that record words related to thinking, perception and expression. To construct phrase patterns, we first split all words from LIWC cognitive processing categories into their part of speech labels (verbs, nouns, adjectives).  Consider the verb \textit{prefer} and noun \textit{preference} from LIWC cognitive processing category. Both words, when paired with different pronouns can form expressions of opinions. For example, \textit{``I (first person subjective) prefer''} and \textit{``My (first person possessive) preference''} both indicate personal opinions. Moreover, variations such as negations (\textit{``I do not prefer''}) or adjectives (\textit{``My strong preference''}) also signal opinions. Hence, we build our initial set of opinion patterns from LIWC's cognitive processing verbs, nouns and adjectives by pairing them with appropriate pronouns and variations. We iteratively improve upon this list of phrase patterns by first, extracting sentences containing those patterns and then, manually eliminating verbs, nouns and adjectives that do not signal opinions. Table \ref{tab:op_pat} in the Appendix lists the final patterns and examples. Note that our opinion extraction method is based on LIWC cognitive processing lexicon that contains limited number of words. Hence, it is possible that our opinion extraction misses out of some expressions of opinions.

    % %Next, we consider different pronouns and their type (subjective and possessive) separately and pair them with appropriate part of speech words. 
    
    % %Next we consider the possibility of negations, auxiliary verbs and adverbs, Based on the initial set of patterns, we identify sentences containing those patterns. 
    % Then, we iteratively improve our candidate list of verbs nouns and adjectives along with the pattern structures by examining the sentences of extracted patterns. .  While, we follow the same methodology as other researchers identifying behaviors such as information seeking \cite{yang2019seekers} and informational and emotional support \cite{biyani2014identifying} note that our opinion extraction system is high precision low recall. Meaning, we do not claim to extract all the patterns present in the text. 
 
    \item \textbf{Expressions of Solicitation in Posts with Links from Extremist Websites: } In solicitation, we want to identify expressions that demand some action on the reader's part. Here, we are looking for calls for donations, invitations for events and protests and participation in policy advocacy (e.g., sign the petition, call your representative etc). To extract solicitation patterns, we follow similar procedure as opinion extraction but instead look at verbs, nouns and adjectives from LIWC's social and affiliation categories. LIWC social and affiliation categories contain words such as \textit{sign}, \textit{call}, \textit{contact}. 
    We build phrase patterns and iteratively evaluate them using methods similar to opinion extraction and report final solicitation phrase patterns in Table \ref{tab:sol_pat} in the Appendix. 
    
    \end{itemize}

\subsection{\textbf{Clustering Extremist Accounts Based on the Derived Features} }
We use the features described above, to cluster the extremist accounts and label each cluster as a role in the extremist movement. 
We use the theory based features (described in the previous subsection) representing drives, engagement and strategies to discriminate between different roles in meaningful way. For role identification, we first need to decide on the number of roles and then use a technique that integrates structural data (feature vectors for extremist accounts) with interpretive analysis that will allow us to describe roles in a relevant way. We use K-Means clustering---an unsupervised clustering algorithm commonly used by other CSCW scholars in role identification studies \cite{arazy2017and}. For example, Arazy et. al. used K-Means to to identify emergent roles in Wikipedia contributors \cite{arazy2017and}. We use the popular K-Means method with kmeans++ initialization to cluster the extremist accounts based on their activity in the six month time window $T_{1}$ (Figure \ref{fig:method} (c)). Specifically, we represent every extremist account with a feature vector of length 13 (6 drives $+$ 5 engagement $+$ 2 strategies). Next, we perform a series of robustness checks to first, determine the number of clusters in order to obtain the optimal separate between different roles and then, to check the stability of clusters. All features were standardized for the downstream analysis.
In K-Means algorithm, the number of clusters , $K$ is a free parameter. We first find the best value of $K$ with an elbow analysis that offers a natural trade-off between the best separation between the clusters and the number of clusters. Specifically, we train the K-Means algorithm for number of clusters ranging from 2 to 20 and plot distortions---sum of squared distances from each point to its assigned center. We observe the elbow at $K=5$ (see Figure \mbox{\ref{fig:elbow} (a)} in the Appendix). Thus we assume 5 as the optimal number of roles. We repeated the elbow experiment with other scoring parameters such as silhouette distance---a measure of how similar a data point is to its own cluster compared to other clusters---with similar results. We also check for the stability of final cluster assignments with various random seeds. Moreover, to check the robustness of our method, we perform the clustering with alternate clustering methods observing similar elbow and cluster assignments. Section \mbox{\ref{sec:cluster_robust}} contains more details about the robustness checks used for the cluster analysis.
We assume that every cluster generated, represents a role in the online extremist movement. 
Next, we use expert guided interpretive analysis to label the roles and their descriptions. 

\subsubsection{\textbf{Role Labeling with Expert Evaluation}}
% We follow a two-fold strategy for role labeling. First we look at the composition of each cluster by observing multidimensional means. Meaning, we look at mean values and distributions of all features for each cluster and highlight the distinguishing features for every cluster. Next, we look at top 5 most representative accounts for every cluster (accounts with the closest distance to their cluster center) and read top 10 most liked posts from every page. Next, together with the distinguishing features and the qualitative analysis of Facebook posts, we label the clusters symbolizing their most representative traits.
Do our quantitatively identified clusters represent coherent roles in extremist movement participation? To evaluate, 
we invited a group of seven social psychology and social movement experts to first analyze and then label the clusters based on their representative characteristics. This group consisted of one senior professor, one assistant professor, one post-graduate researcher and four senior doctoral students. 
To reduce bias in role labeling, we selected the external evaluators who are not a part of the author group and were not involved in any of the work preceding or following this stage. We showed them mean feature values for every cluster. Additionally, we selected top 5 representative extremist accounts from each cluster---accounts with closest distance to the cluster centers. We compiled the list of top 10 most representative posts from each selected account ranked by post likes \cite{horne2017identifying}. Based on this information, we asked the evaluators to come up with labels and descriptions for each cluster. In all, every evaluator looked at 250 Facebook posts (5 clusters $\times$ 5 accounts $\times$ 10 posts) and recorded the possible labels and descriptions. Finally, the first author and the evaluators together, worked and selected the best label for every cluster. We present the identified roles and comments on the expert evaluation process in the next section.

\section{\textbf{RQ1 Results: Roles in Online Extremist Movements}}

%\subsection{\textbf{Roles in Extremist Information Mobilization}}

% \usepackage{booktabs}
% \usepackage{graphicx}
\begin{table*}[t]
\centering
\sffamily
\resizebox{\textwidth}{!}{%https://www.overleaf.com/project/5e8899176517890001543447
\begin{tabular}{llll}
\hline
\textbf{Role} & \textbf{Frequency} & \textbf{Example texts used by the accounts while sharing links from extremist websites} & \textbf{} \\ \hline
\textbf{Solicitors} & 5.2\% & \textit{\begin{tabular}[c]{@{}l@{}}"...Sign here to demand her {[}Rep. Maxine Waters{]} immediate resignation"\\ "...Join us in signing thank you card for President Trump"\\ "...Stand with us to take back our streets"\end{tabular}} & \textit{} \\ \hline
\textbf{Educators} & 10.6\% & \textit{\begin{tabular}[c]{@{}l@{}}"...Escaping from motherhood: how it destroys society"\\ "...We believe that we have the duty to instruct people in the truth of Tradition. Even if it destroys their party"\\ "...The need for an ethnocentric society amidst "globalism" "\end{tabular}} & \textit{} \\ \hline
\textbf{Flamers} & 18.4\% & \textit{\begin{tabular}[c]{@{}l@{}}"...genuine Christians know that homosexuality is an abomination before GOD! "\\ "...MURDERED in cold blood. Emergency: Gunfire, bodies and BLM murderers"\\ "...house democrats vote to allow female genital mutilation..."  (false information flag by Facebook pops up)\end{tabular}} & \textit{} \\ \hline
\textbf{Motivators} & 29.4\% & \textit{\begin{tabular}[c]{@{}l@{}}"...Senator Dan Halls stands with us in a passionate commitment to strengthening religious freedom"\\ "...FREE SPEECH WINS!!! Supreme court rules pregnancy centers can't be forced to advertise abortion"\\ "...we WILL have the COURAGE to defend ourselves! borders, language and culture MATTER"\end{tabular}} &  \\ \hline
\textbf{Sympathizers} & 36.4\% & \textit{\begin{tabular}[c]{@{}l@{}}"...White South Africans petition Trump to allow them to migrate to the US"\\ "...A jihadi cult member running for Congress as Democrat from Alaska"\\ "...They will only see Italy on postcard: Italy turns away another migrant ship"\end{tabular}} & \\ \hline
\end{tabular}%
}
\caption{Roles and the corresponding percent of extremist accounts in the dataset. The examples are of the texts written by the extremist accounts while sharing links from the extremist websites}
\label{tab:roles}
\end{table*}

Here we describe the roles played by extremist accounts and their typical behaviors. Table \ref{tab:roles} displays the frequency of the roles in the dataset alongwith the example text written by the extremist accounts while sharing links from extremist websites. We identified five roles in the online extremist movements. 

\begin{enumerate}
    \item \textbf{Solicitors: } These are the accounts that solicit participation from their readers for signing petitions, attending rallies etc. On average, around 20\% of their links come extremist domains and they post extremist content with fairly consistent trend throughout the six month period (trend feature values close to 0). These accounts use high group identity language such as ``we'', ``our'', ``us'' compared to other roles. Evaluators also described them as ``recruiters.'' One evaluator mentioned: 
    \begin{quote}
    \textit{``These groups appear to be soliciting action for their hate. To some extent, they seem pretty keen on doing something about the groups they hate and are actively sharing/liking posts to promote action''}
        
    \end{quote}

    \vspace{2pt}
    \item\textbf{Educators: } Educators have distinctively high amount of extremist content in their link sharing. On an average, 50\% of their links come from extremist domains. Additionally, the extremist links posts get more likes and comments compared to other material on these pages/groups. They post the extremist content with consistently high rates (trend feature > 0) throughout the six months. In qualitative evaluation, the experts pointed out that these groups share intellectual material and appear serious and sincere in propagating the fundamentals of extremist ideologies. The evaluators also suggested alternate labels such as ``preachers'' and ``intellectuals.'' According to one evaluator:
    
    \begin{quote}
        \textit{``...they seem to take effort to make logical arguments. They are not necessarily showing anger towards  other groups but are instead more focused on highlighting their own group's worth logically/analytically''}
    \end{quote}
    \vspace{2pt}
    
    \item \textbf{Flamers: } These accounts spew toxic and inflammatory content. Around 5\% of their links belong to extremist domains and the messages on the links and the link text itself often contains language suggesting anger and injustice. In other words, these pages/groups have the highest proportion of anger and injustice related words while disseminating extremist content. The extremist links posted on these accounts get higher number of shares compared to the rest of the content. The experts also described them them as ``fear mongerers'' for attempting to cause general outrage. Immediately after looking through the posts, one evaluator commented:
    \begin{quote}
        \textit{``these are clearly very strong, divisive and toxic posts''}
    \end{quote}
    \vspace{2pt}

    \item \textbf{Motivators: } Around 7\% of the links by motivators are sourced from extremist domains. Evaluators pointed out that motivators use exceptionally positive language. While posting extremist content, they stress on the achievements and rewards associated with extremist activities. Motivators also express opinions with highest proportions compared to the other roles. Experts noted that these accounts engage in policy activism focusing on policies protecting and defending cultural and moral values. Evaluators also mentioned: 
    \begin{quote}
        \textit{``it almost looks like they are celebrating the in-group [people and organization involved in the extremist movement] and the sensationalized news about the in-group''}
    \end{quote}
    \vspace{2pt}
    
    \item \textbf{Sympathizers: } These accounts post extremist content links with lowest rates (2\% of their Facebook link posts) and sporadically throughout the six month period. They also show low engagement in terms of likes, shares and comments on the extremist link posts. According to the experts, these groups are on the fringe of extremist ideology and might be only slightly interested in extremist causes. Experts also referred to them as ``observers.'' One evaluator described \textit{sympathizers} as: \textit{``They look more like general conservative interest groups''}
\end{enumerate}

% Table \ref{tab:roles} lists the roles alongwith the percent accounts and examples of link posts from the roles. 
\vspace{3pt}
\noindent{\textbf{Notes on Expert Evaluation:} } Manually evaluating and labeling roles is a challenging but an insightful task. Interpreting clusters and finalizing the role labels  was an iterative, discursive process. Some roles were easy to identify and label (for example, \textit{solicitors}, \textit{educators} and \textit{flamers}) while others required going over additional details such as page/group names and their descriptions. For example, while the experts reached at immediate consensus for the \textit{flamers} role, they deliberated over the \textit{motivators} category.  
However, qualitative evaluation provided additional insights about the clusters that were not apparent from the original feature set. For example, we found that accounts in the \textit{flamers} category had higher amount of posts flagged as misinformation or violent/graphic content. Similarly, \textit{educators} shared links containing material describing extremist philosophies (for example, white identity, ethnocentrism, political philosophy) than rest of the accounts. Note that such nuances are harder to capture computationally and further extensive qualitative study of these pages might provide new theoretical insights about actors in online extremist movements.

\section{\textbf{RQ2 Method: Measuring Role Dynamics}}
\begin{figure*}[t]
    \centering
    \includegraphics[width=0.99\textwidth]{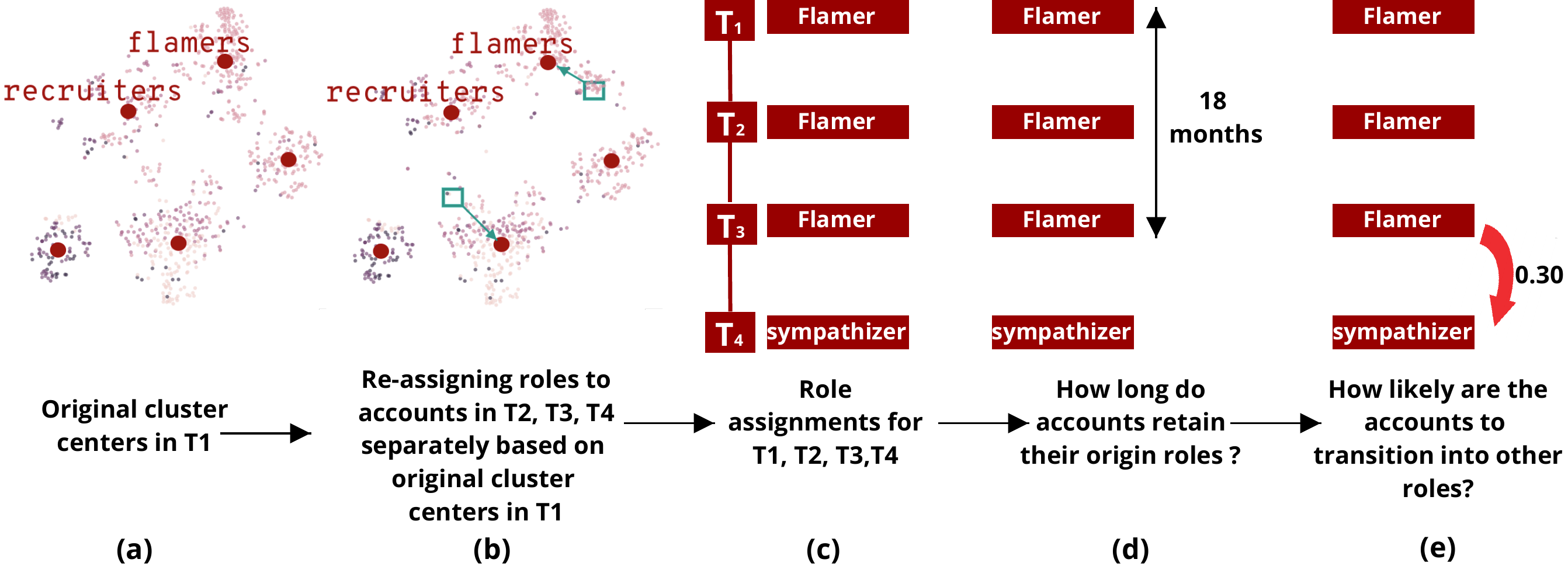}
    \caption{Figure showing method steps for analyzing role dynamics. (a) We identified cluster centers for each role in RQ1 using the account activity in $T_1$ (Jan 2018-Jun 2018). (b) We use those cluster centers identified in $T_{1}$ and re-assign extremist accounts to roles based on their activity in $T_{2}$, $T_{3}$ and $T_{4}$. (c) For every account in the dataset, we get cluster assignments for all time periods. (d) To measure role retention, we calculate the number of time periods for which the extremist accounts maintain their original roles from $T_{1}$. (e) In role transition, we calculate the probability with which the extremist accounts may transition to different roles.}
    \label{fig:rq2_method}
\end{figure*}

The roles we identify are data-driven. The underlying characteristics of the roles are derived mainly from the theoretical studies on physical protest events. Unlike physical social movements where members can commit to protests, meetings or other events, social media provides a dynamic, evolving space for members to engage or disengage from the social movement as they like without much accountability \cite{corrigall2011patterns}. Hence, we also analyze the dynamics of roles based on how long the accounts retain their initial roles (role retention) and their transition probability to another role (role transition). Figure \ref{fig:rq2_method} displays the method steps taken to measure role retention and role transition. 

\vspace{3pt}
\noindent \textbf{RQ2a: Measuring Role Retention: }
By retention, we measure how long the extremist accounts adhere to their originally identified roles. Recall that our data spanning two years (Jan'18 to Dec'19) is sliced into four windows, each six months duration---$T_{1}$, $T_{2}$, $T_{3}$, $T_{4}$ (Figure \ref{fig:method} (b)). We initially identified roles in RQ1 using the account activity in the first time period, $T_{1}$ (Jan 2018 - Jun 2018). To measure role retention, we use the cluster centers (mean values for each feature for each cluster) from $T_{1}$ and re-assign roles to the extremist accounts using their activity in $T_{2}$, $T_{3}$, $T_{4}$. 
For every extremist account, we now have the role assignment for $T_{1}$, $T_{2}$, $T_{3}$, and $T_{4}$. 
For example in Figure \ref{fig:rq2_method} (c), an account stays in the \textit{flamer} role for three consecutive time periods. That is, the role identified in  $T_{1}$ is retained for (3 X 6) 18 months consecutive months. Note that all time windows $T_{1}$, $T_{2}$, $T_{3}$, and $T_{4}$ represent distinct calendar months between 2018 and 2019 as displayed in Figure \mbox{\ref{fig:method}} (b). Using the initially defined cluster centers in $T_{1}$ allows us to compare the accounts' activities in $T_{2}$, $T_{3}$, $T_{4}$ with respect to their own past states. 
%Note that all the features used in clustering are normalized by the overall activity of the account. Hence, the external factors such as natural increase or decrease in the accounts' overall link posting rate on Facebook should not affect the re-clustering in the later time windows. 
How long do extremist accounts adhere to their initially identified role in $T_{1}$ over the subsequent future time windows? To answer, we calculate the number of continuous time windows across which an account retains its initially identified role. Finally, for each role, we calculate the proportion of accounts maintain their roles across just one ($T_{1}$), two ($T_{1} \rightarrow T_{2}$), three ($T_{1} \rightarrow T_{3}$) or all four ($T_{1} \rightarrow T_{4}$) time windows. 

\vspace{3pt}
\noindent \textbf{RQ2b: Measuring Role Transition: }
How likely are the extremist accounts to move from one role to another? For example, how likely are \textit{sympathizers}---fringe supporters of the extremist movement---to transition to \textit{educators}---accounts that actively distribute large proportion of extremist content? From the role retention analysis, we know which role every extremist account plays in each of the time windows---$T_{1}$, $T_{2}$, $T_{3}$, $T_{4}$. For every account, we consider the role played by that account in a particular time window $T_{i}$ as the \textit{state} $S_{i}$ that account is in. Consequently, for every account we have sequence of four states corresponding to each of the time windows. The example account in Figure \ref{fig:rq2_method} (e) has states: \textit{flamer} $\rightarrow$ \textit{flamer} $\rightarrow$ \textit{flamer} $\rightarrow$ \textit{sympathizer}. 
Using such state sequences of all extremist accounts, we then calculate the state transition, or the role transition probability. Specifically, we calculate the pairwise transition probabilities for each pairs of roles. 
High probability of transition \textit{solicitor} $\longrightarrow$ \textit{educator} will indicate that an account currently playing the role of \textit{solicitor} is likely to transition into \textit{educator} in the next time window with high probability.   

%  We train the K-Means model and assign clusters to every account using their activity in  $T_{1}$. Using the cluster centers from $T_{1}$, we predict the roles for each account for the successive time buckets. Prediction using initially defined cluster centers allows us to compare the accounts' activity with respect to their own past state. Note that all clustering features are normalized by the overall activity of the account. Hence, the external factors such as natural increase or decrease in the posting rate should not affect the cluster predictions in the later time buckets. Finally, for every account, we have cluster assignments for $T_{1}$, $T_{2}$, $T_{3}$. First to measure the role retention state, we simply calculate the number of continuous time buckets for which the account retains its initially identified role. Next, to understand role transition, we calculate the transition probabilities of roles based on the sequence of roles obtained across the time buckets. We discuss the identified roles and their definition alongwith the role retention and transition results in the next section. 

\section{\textbf{RQ2 Results: Role Dynamics}}
%How stable are these roles over time? Here we present our results for role retention and transition analysis.  

\begin{figure*}[t!]
    \centering
    \begin{subfigure}[t]{0.45\textwidth}
        \centering
        \includegraphics[width=0.70\textwidth]{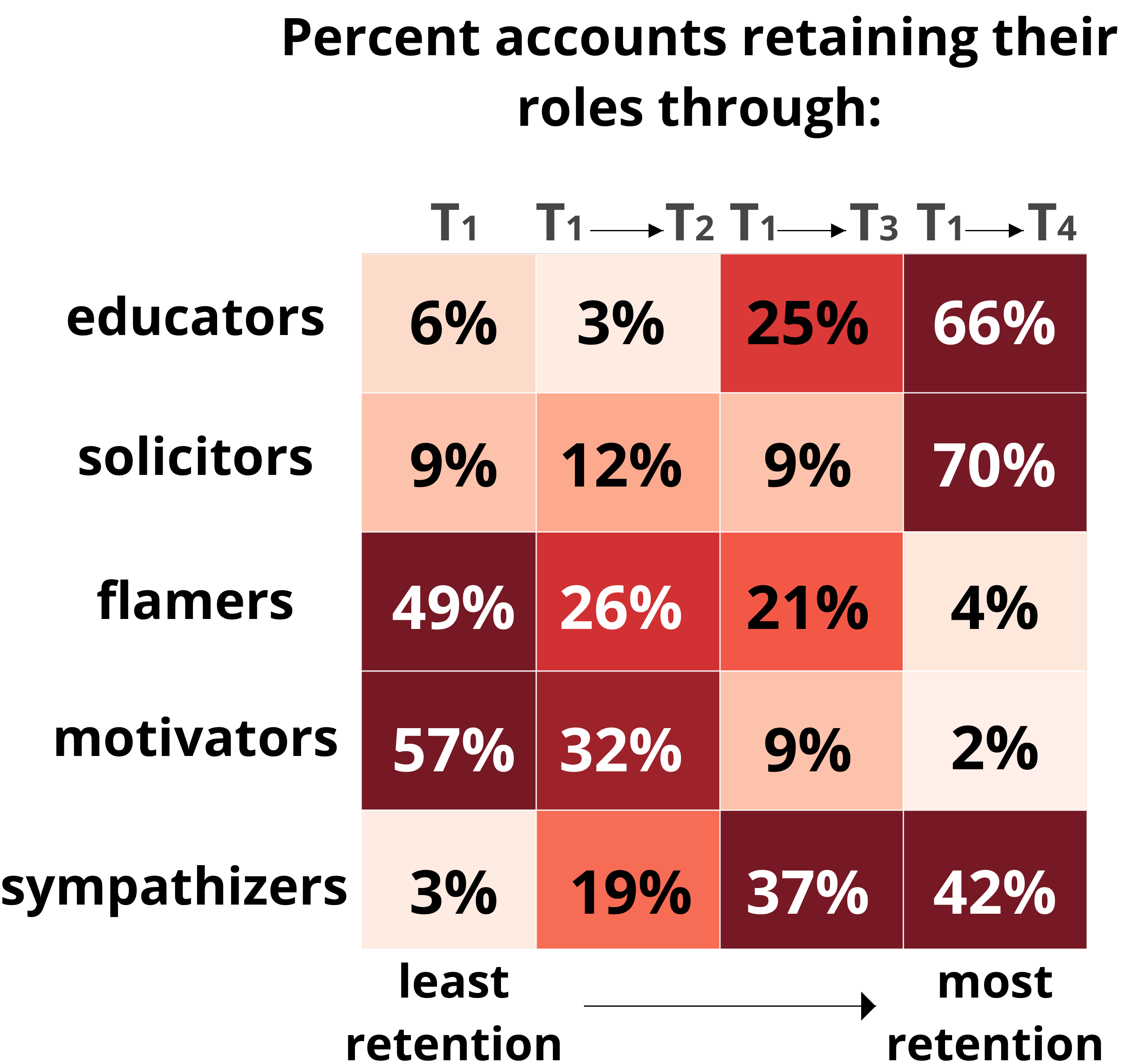}
        \caption{}
    \end{subfigure}%
    ~ 
    \begin{subfigure}[t]{0.45\textwidth}
        \centering
        \includegraphics[width=0.70\textwidth]{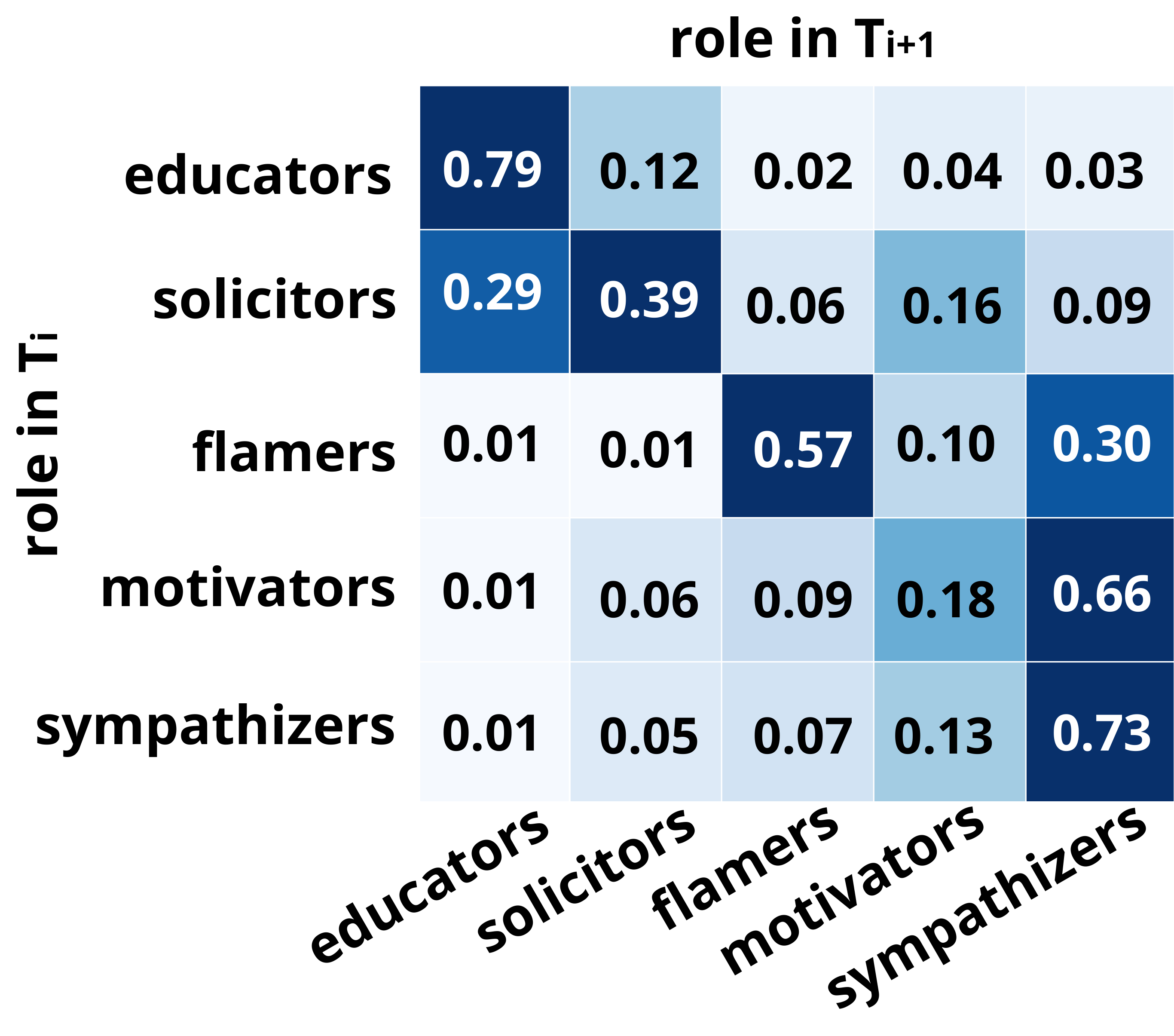}
        \caption{}
    \end{subfigure}

    \caption{Figure presenting the results of RQ2. (a) indicates role retention---proportion of extremist accounts that retain their originally identified role through subsequent time windows in the dataset. For example, 70\% of the \textit{solicitors} retain their role throughout $T_{1} \rightarrow T_{4}$. Whereas 49\% of \textit{flamers} retain their roles for only the initial $T_{1}$ period. (b) displays the role transition probability matrix. Rows indicate the role in $T_{i}$ and columns indicate the role in $T_{i+1}$. The cells indicate the probability of transition from role in $T_{i}$ to $T_{i+1}$. For example, \textit{solicitors} may transition to \textit{educators} in the next time period with 0.29 probability.}
    \label{fig:stability}
\end{figure*}

\noindent \textbf{RQ2a: Retention:} Figure \ref{fig:stability} (a) displays the initial roles (rows) and number of time periods ($T_{1}$, $T_{2}$, $T_{3}$, $T_{4}$) for which the role was retained by an account. We find that 66\% of the \textit{educators} and 70\% of the \textit{solicitors} retain their initial role for all four time periods, that is for the entire two years. Whereas, 49\% of the \textit{flamers} and 57\% of \textit{motivators} transition to another role just after $T_{1}$ (6 months). \textit{Educators} and \textit{solicitors} can be viewed as an elite group in extremist movements---members who distribute (information) resources and actively recruit others \cite{MccarthyResourceTheory}. On the other hand, \textit{flamers} and \textit{motivators} are supporters who exhibit low engagement with the links from extremist websites. Our results suggest that roles more core to the extremist movement such as \textit{educators} and \textit{solicitors}) are more stable. In other words, \textit{educators} and \textit{solicitors}) are more likely to maintain their roles and consequently their behaviour surrounding the participation in extremist movements for longer periods compared to others. 

% \begin{figure*}[t]
%     \centering
%     \includegraphics[width=0.30\textwidth]{figures/rq2_transition_final.pdf}
%     \caption{Role transition probabilities}
%     \label{fig:transition}
% \end{figure*}

\vspace{3pt}
\noindent \textbf{RQ2b: Transition:} Figure \ref{fig:stability} (b) displays a transition matrix for roles. The values in the cell indicate the probability by which an account in one role (row) would transition to another (column) in the next six months. Based on the main diagonal, accounts in most role are more likely to retain the same role in the next time window. For example, \textit{educators} will stay \textit{educators} in the next six months with 0.79 probability. Similarly the probability of \textit{flamer} $\longrightarrow$ \textit{flamer} is 0.57. Notably, \textit{solicitors} and \textit{educators}---roles with highest engagement with links from extremist websites---have highest transition probabilities with each other compared to any other roles (\textit{solicitor} $\longrightarrow$ \textit{educator} $=0.29$ and \textit{educator} $\longrightarrow$ \textit{solicitor} $=0.12$). Moreover, \textit{flamers} and \textit{motivators} can transition to \textit{sympathizers} with $0.30$ and $0.66$ probabilities respectively, indicating that roles with less engagement with extremist content are also less stable.

 \section{\textbf{RQ3 Method: Influential Roles in Information Sharing }}

\begin{figure*}[t]
    \centering
    \includegraphics[width=0.90\textwidth]{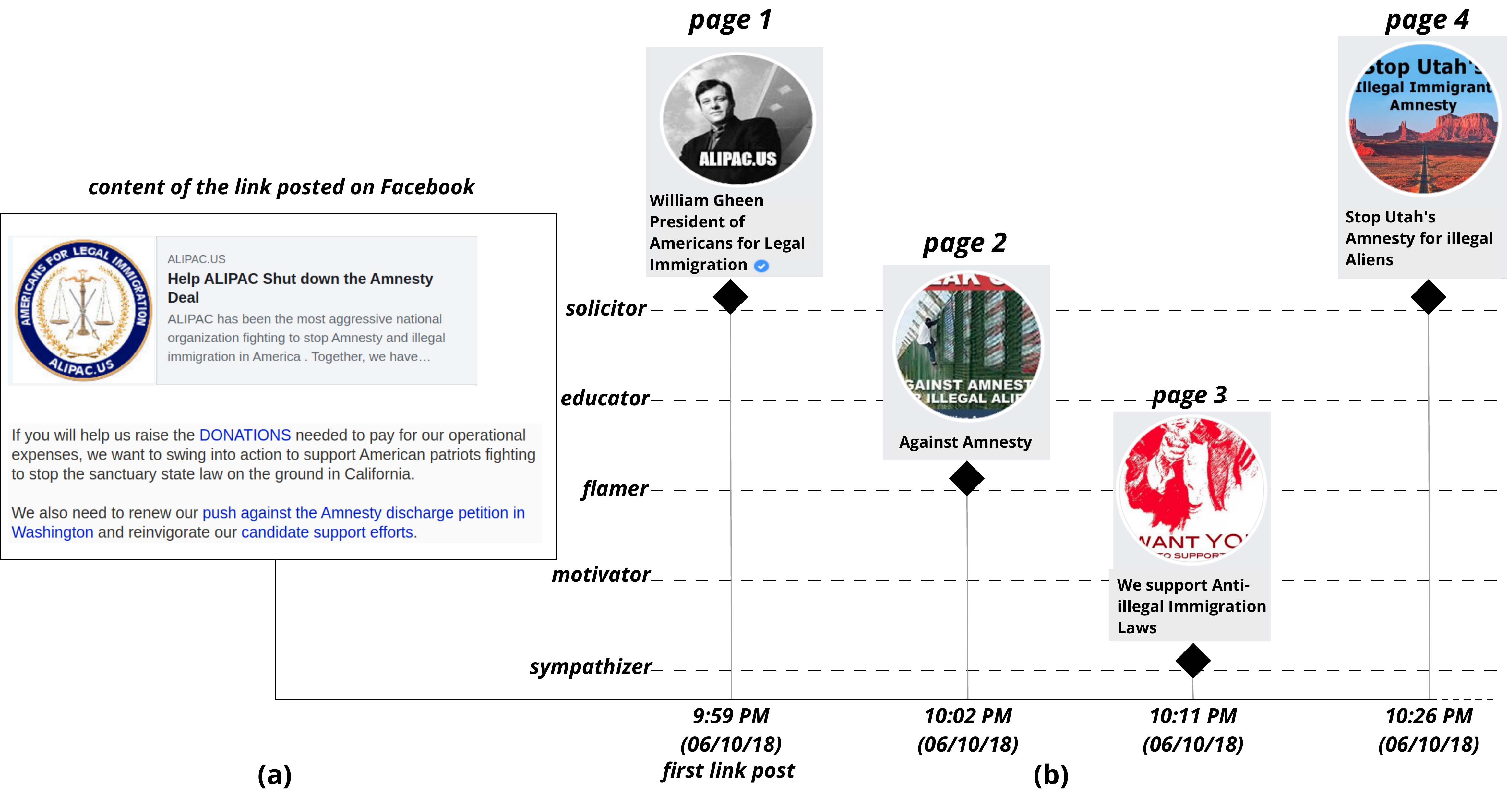}
    \caption{An example illustrating a series of link posting events by accounts playing various roles (a) Contents of the link hosted on the website of an SPLC designated anti-Immigration group ALIPAC. This link contains call for action against stopping immigrant amnesty deal in California (b) series of link posting \textit{events} on Facebook demonstrating how a link from an extremist website flows through various accounts playing roles in extremist movements. The link was first posted on a Facebook page of the president of the ALIPAC (page 1). Next, the link was posted by various accounts---Against Amnesty (page 2), We support Anti-illegal Immigration Laws (page 3) and Stop Utah's Amnesty for illegal Aliens (page 4)---playing various roles. 
    This example demonstrates  a series of 4  events---link posts by \textit{solicitor}, \textit{flamer}, \textit{sympathizer} and \textit{solicitor} accounts.}
    \label{fig:RQ2_motiv}
\end{figure*}

In RQ1 we identified five roles in extremist movements---\textit{educators}, \textit{solicitors}, \textit{flamers}, \textit{motivators} and \textit{sympathizers}. In RQ2, we measured the dynamics of the roles over time. How influential are these roles in disseminating information? Previous research suggests that terrorist organizations such as ISIS are able to mobilize information through series of online accounts that knowingly or unknowingly play a role in advancing terrorist propaganda \cite{krona20195}. Researchers observed a similar pattern of  participatory information sharing in political disinformation campaigns \cite{starbird2019disinformation}. In this research question, we model the information sharing by various roles to investigate how influential different roles are in spreading extremist content, fake news, biased news and conspiracy sources. We open up by a motivating example describing how information crucial to the extremist causes flows through various roles online. 
Next we describe the method details for modeling the information flow using Hawkes process and inferring influence.

\subsection{\textbf{Call for Shutting Down the Amnesty Deal: A Motivating Example}}
On June 10th 2018, ALIPAC---an anti-Immigration political action organization \cite{alipac}---put out a call for action to stop amnesty deal for immigrants in California (Figure \ref{fig:RQ2_motiv} (a)). This included call for donations, and participation to help support ALIPAC's operational costs. A link containing this call for action was posted on Facebook on the same day at 9:59 PM by a verified page managed by the president of ALIPAC (page 1 in Figure \ref{fig:RQ2_motiv} (b)). This Facebook verified page is also known for posting fake news from websites hosting plagiarized content \cite{alipacfake}. According to our RQ1 analysis, this page (page 1) plays the role of \textit{solicitor}. Three minutes after the initial post by page 1, another Facebook page---identified as \textit{flamer}---posted the same link. Following the post by page 2, a \textit{sympathizer} page (fringe supporter of the extremist content) as well as another \textit{recruiter} page also posted the same link. Following these four link posts, another 32 pages posted the same link containing the call for action by an anti-Immigration group over the period of next three days. How influential was page 1 in spreading this link containing the call for action? How influential are \textit{recruiters} or other roles in spreading the extremist content or other types of information? In the next section, we explain our information flow model that considers both temporal order of link posts and the time difference between consecutive link posts to statistically establish the influence of different roles in spreading links from various types of sources.

\subsection{\textbf{Intuitive Background for Measuring Influence in Link Sharing}}
Our primary goal here is to understand how influential roles are in spreading information from various types of sources. By \textbf{influence}, we measure the probability by which a link posting by one role affects link posting by other roles in future. To measure the influence of various roles, we use multivariate Hawkes process which is commonly used for modelling events on social media \cite{goel2016social,zannettou2017web}. For example, Zannettou et. al. used Hawkes process to understand the influence of various online platforms (Reddit, Twitter, 4chan) on each other in information sharing \cite{zannettou2017web}. Hawkes processes are popularly used as a more sophisticated model for measuring influence compared to simple point processes \cite{rizoiu2017tutorial,embrechts2011multivariate}. 
%The popularity of Hawkes process in studying influence structures can be attributed to  its rigorous modelling of nuances in information sharing compared to simple point processes. 
Specifically, Hawkes process can account for various nuances in information sharing. For example, while calculating influence, Hawkes process considers the temporal flow of link sharing, the time between consecutive link posts and the natural link posting tendencies of various roles involved. 
In its simplest form, modeling link sharing by different roles using Hawkes process can provide us with two parameters: (1) the natural link sharing tendency of the roles or \textbf{background rate} and (2) the pairwise influence that roles exert on each other in link sharing. 

More formally, consider an \emph{\textbf{event}} where a \textit{solicitor} posts a link. This event can affect the probability by which other roles (for example, \textit{symapathizers}) also post the same link in future. In other words, link posting by one role might increase the probability of the link posting by another role. The effect of previous events on the current event is additive and decaying. This means that, as more time passes between two events, the lesser impact the previous events will have on the current event. For each link, we extract such \textbf{series of events} (sequence of timestamps indicating the link posting by extremist accounts in various roles). Next, we provide the details for fitting Hawkes process on the extracted series of events.

\subsection{\textbf{Fitting Hawkes Process on Link Posting Events}}
A multivariate Hawkes model consists of K processes each with a background rate of $\lambda_{0,k}$. Fitting a Hawkes model on link sharing by five roles will give us the background link posting rates for each role---probability of an event cased by external factors---as well as the probabilistic estimate of the effect of the roles on each other. We use a discrete-time Hawkes model which is commonly used for modeling information flow on social media \cite{zannettou2017web}. For each link, the time of posting is divided into discrete time bins of duration \textbf{$\Delta t$}, creating a series of events with time granularity of $\Delta t$. 
Fitting Hawkes model on this series of events will give us estimates for the background rate of $\lambda_{0,k}$ and pairwise weight matrix $W^{K \times K}$. Appendix \ref{sec:math_hawkes} provides the formal mathematical representation of Hawkes process along with its parameters. 

% This series is a Hawkes process if the rate of each process has the parameterized form:

% \begin{equation}
%     \lambda_{t,k} = \lambda_{0,k} + \sum_{k^{'}=1}^{K} \sum_{t^{'}=1}^{t-1} s_{t^{'},k^{'}} \cdot h_{k^{'}\rightarrow k}[t - t^{'}]
% \end{equation}

% where $\lambda_{0,k}$ is the background rate, $s$ is the matrix of events generated from common link sharing and $h$ is an impulse response function describing the amplitude of influence that events on process (or role) $k^{'}$ have on the rate of process (or role) $k$. Guided by the descriptions in \cite{zannettou2017web,linderman2014discovering}, we can further decompose $h$ into the weight matrix $W$ and probability mass function $G$. 

For our analysis, the \textbf{weight matrix $W^{K \times K}$} is most relevant. Every value $w_{i,j}$ in the weight matrix $W^{K \times K}$ signifies the expected number of subsequent events that will be caused in the process $j$ after an event in the process $i$. In other words, value in weight matrix $W_{\textit{i} \rightarrow \textit{j}}$ indicates the strength with which a link posting event by role $i$ can affect link posting by role $j$ \cite{zannettou2017web}. 
We consider these weights as a proxy for \textbf{influence} of role $i$ on role $j$. 
Note that weights from $W$ describe the expectations of event occurrence above the background rate $\lambda_{0,k}$ of the process where background rate $\lambda_{0,k}$ accounts for events due to external factors.

\subsection{\textbf{Measuring Influence in Link Sharing}} 
To understand the influence of roles, we consider the link posts made by the extremist accounts in $T_{1}$ (Jan 2018-Jun 2018) and consider the role identified for every extremist account in RQ1. We assess the influence of roles by using the weight matrix $W$ described in the previous subsection. To calculate the weight matrix, we first need to determine the number of processes ($K$) and the length of discrete time windows $\Delta t$ to bin every link  posting event. Since we aim to examine how influential various roles are, we model the link postings with K=5 processes---one for every role identified in RQ1. To select $\Delta t$, we plotted the distribution of inter arrival times---time in seconds between two consecutive link posting events---for all links. The goal for choosing $\Delta t$ is to have a separate bin for most events.  We decided $\Delta t=30 sec$ based on the $10^{th}$ percentile cut-off of the distribution. Choosing this percentile cut-off allows us to put approximately 91\% of the link posting events in the bin by themselves. To eliminate links that are shared less, we select links that are shared by atleast 10 different extremist accounts spanning over atleast 3 of the 5 identified roles. Next, we fit a Hawkes model for each link using a nonparametric Expectation Maximization (EM) for parameter estimation \cite{lewis2011nonparametric}. With the EM algorithm we are able to get the background rate ($\lambda_{0,k}$) as well as the weight matrix $W$ described in the previous subsection. Both, background rates for each role and the weight matrix, contain non-negative real values. We further row normalize the weight matrix to represent the influence of a role in a row on the role in a column with values bound between 0 to 1.
Note that so far we have calculated the weight matrix $W$ for each link separately. In other words, for each unique link, we obtain pairwise influence values for roles. Remember that in Section \ref{sec:sourcetype}, we recorded link domain types as extremist, biased, fake or conspiratorial based on the OpenSources annotations. Here, we calculate the overall weight matrix for each domain type by averaging the weight matrices of links based on their domain types. For example, to obtain the overall influence of roles in spreading links from fake news domains, we consider all links that are generated from fake news domains and calculate the average of their weight matrices. In the next section, we discuss our results and influential roles in spreading links from extremist, biased, fake news and conspiracy domains. 

% Finally, to understand the overall importance of roles, we calculate the mean of weight matrices obtained from fitting Hawkes process on each link in the dataset grouped by the source type---hate content, biased news, fake news and conspiracy. 

% \subsection{\textbf{Modeling Information Flow for Various Information Sources}}
% Previous researchers have found the evidence of mis and disinformation sources in online extremist discourse \cite{Starbird2018EcosystemDomains,Phadkecross}. Here we try to understand how important each role is in spreading links from hate, fake news, biased and conspiracy/pseudoscience sources. First, we identify the hate, fake news, biased news and conspiracy domains using the dataset published by OpenSources \footnote{\link{https://github.com/BigMcLargeHuge/opensources}}. Table \ref{tab:event_table} lists the number of domains in each source type. Next, we extract link domains from every link in the dataset and group links based on their source type as hate, fake news source, biased source and conspiracy/pseudoscience source. Finally, we model links posting events using the steps described in previous sections, through various roles for each type of source. In the next section, we discuss the results separately for each source type.  

% \usepackage{booktabs}
% \usepackage{graphicx}
\begin{table}[]
\centering
\scriptsize
\resizebox{\textwidth}{!}{%
\begin{tabular}{@{}llllllllll@{}}
\toprule
\multicolumn{1}{c}{\textbf{source type}} & \multicolumn{1}{c}{\textbf{\begin{tabular}[c]{@{}c@{}}\#domains \\ labeled\end{tabular}}} & \multicolumn{1}{c}{\textbf{\begin{tabular}[c]{@{}c@{}}\#labeled domains \\ present in $T_{1}$\end{tabular}}} & \multicolumn{1}{c}{\textbf{\begin{tabular}[c]{@{}c@{}}\#unique \\ links\end{tabular}}} & \multicolumn{1}{c}{\textbf{\begin{tabular}[c]{@{}c@{}}\#events \\ (link posts)\end{tabular}}} & \multicolumn{5}{c}{\textbf{\% link posts made by}} \\ \midrule
 & \textbf{} & \textbf{} & \textbf{} & \textbf{} & \textbf{solicitors} & \textbf{educators} & \textbf{flamers} & \textbf{motivators} & \textbf{sympathizers} \\ \midrule
\textbf{extremist} & 289 & 231 & 758 & 16,532 & 12\% & 28\% & 9\% & 5\% & 44\% \\
\textbf{biased} & 133 & 94 & 1279 & 13,290 & 11 & 9\% & 19\% & 23\% & 36\% \\
\textbf{fake} & 304 & 107 & 380 & 3583 & 9\% & 20\% & 23\% & 13\% & 34\% \\
\textbf{conspiracy} & 154 & 68 & 936 & 10,001 & 20\% & 20\% & 10\% & 21\% & 29\% \\ \bottomrule
\end{tabular}%
}
\caption{Table describing link posting activities by various roles. We first report the number of labeled domains in each category---extremist, biased, fake and conspiracy and also specify how many of those domains are present in the link posts made in $T_{1}$. Next, we list the number of unique links and the number of link posting events for each source type.
%We split the links by their source types---extremist, biased, fake and conspiracy. For each source type, we list number of unique links and link domains. Next, we report total number of \textit{events} for a source type---number of times links from that source type were shared.
We also report what percent of such link posting events were made by various roles. For example, \textit{solicitors} contribute to 12\% of the link posts from the extremist domains. } 
\label{tab:event_table}
\end{table}

\section{\textbf{RQ3 Results: Influential Roles in Information Sharing }}
We summarize the number of links and number of link posting events generated for each type of information in Table \ref{tab:event_table}. Looking at the number of events, \textit{sympathizers} make up largest percent of link posting events in all source types. This is not surprising given that 36.4\% of the accounts are \textit{sympathizers}. Along the same lines, \textit{solicitors}---who actively solicit participation by posting extremist links and \textit{educators}---who share largest proportion of extremist links---contribute to high percent of events in posting extremist links. Interestingly, \textit{flamers}---accounts that often post inflammatory and violent content---are also second highest in posting links from the fake news sources. Moreover, \textit{motivators}---who focus on efficacy of policy changes and use opinionated language, make up 23\% of the biased news posting events. Interestingly, \textit{solicitors}, \textit{educators}, \textit{motivators} and \textit{sympathizers}, all post links from conspiracy sources with similar rates. 

\begin{figure*}[t!]
    \centering
    \begin{subfigure}[t]{0.4\textwidth}
        \centering
        \includegraphics[width=0.90\textwidth]{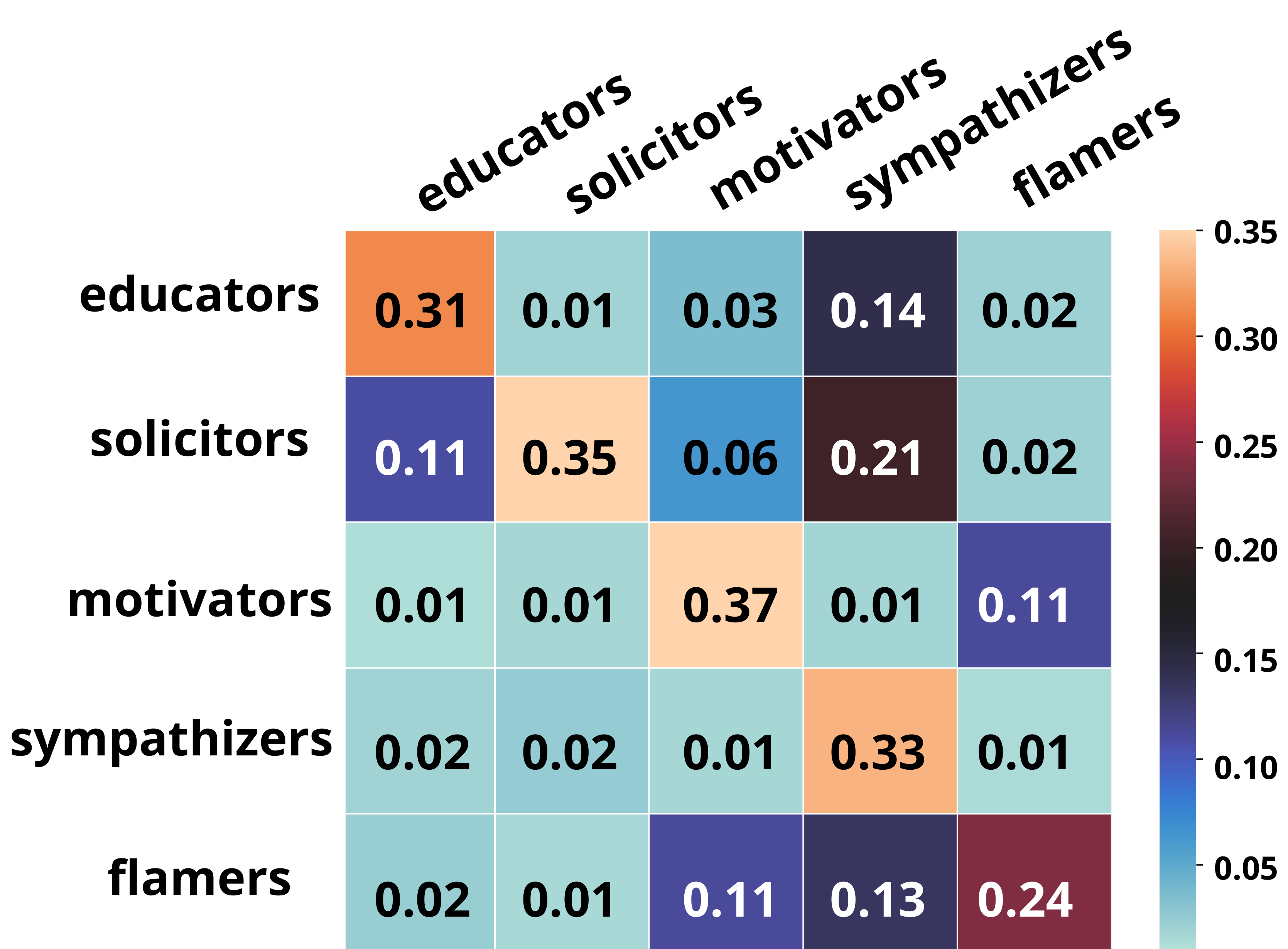}
        \caption{Extremist sources}
    \end{subfigure}%
    ~ 
    \begin{subfigure}[t]{0.4\textwidth}
        \centering
        \includegraphics[width=0.90\textwidth]{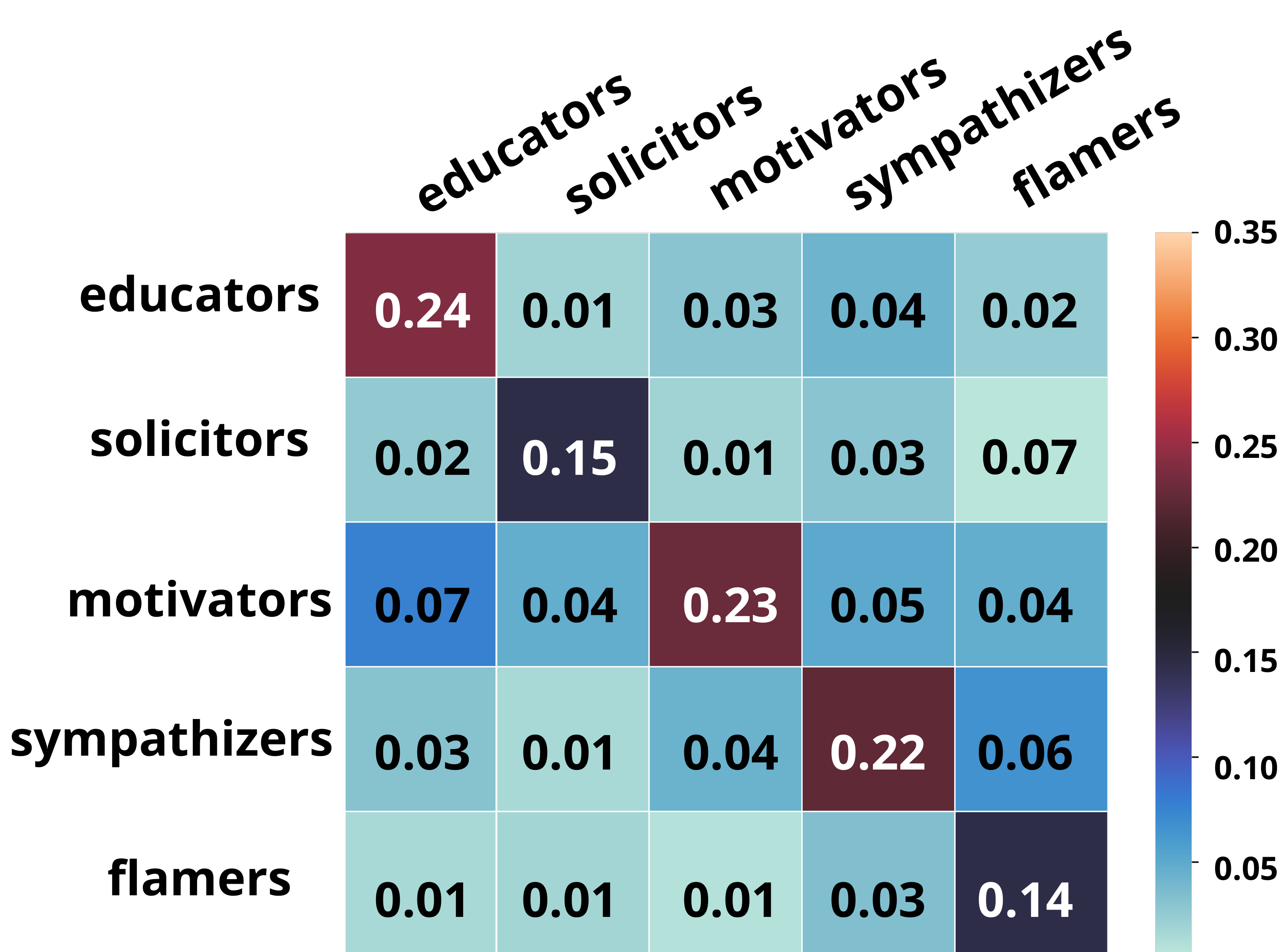}
        \caption{Biased sources}
    \end{subfigure}
    
    \begin{subfigure}[t]{0.4\textwidth}
        \centering
        \includegraphics[width=0.90\textwidth]{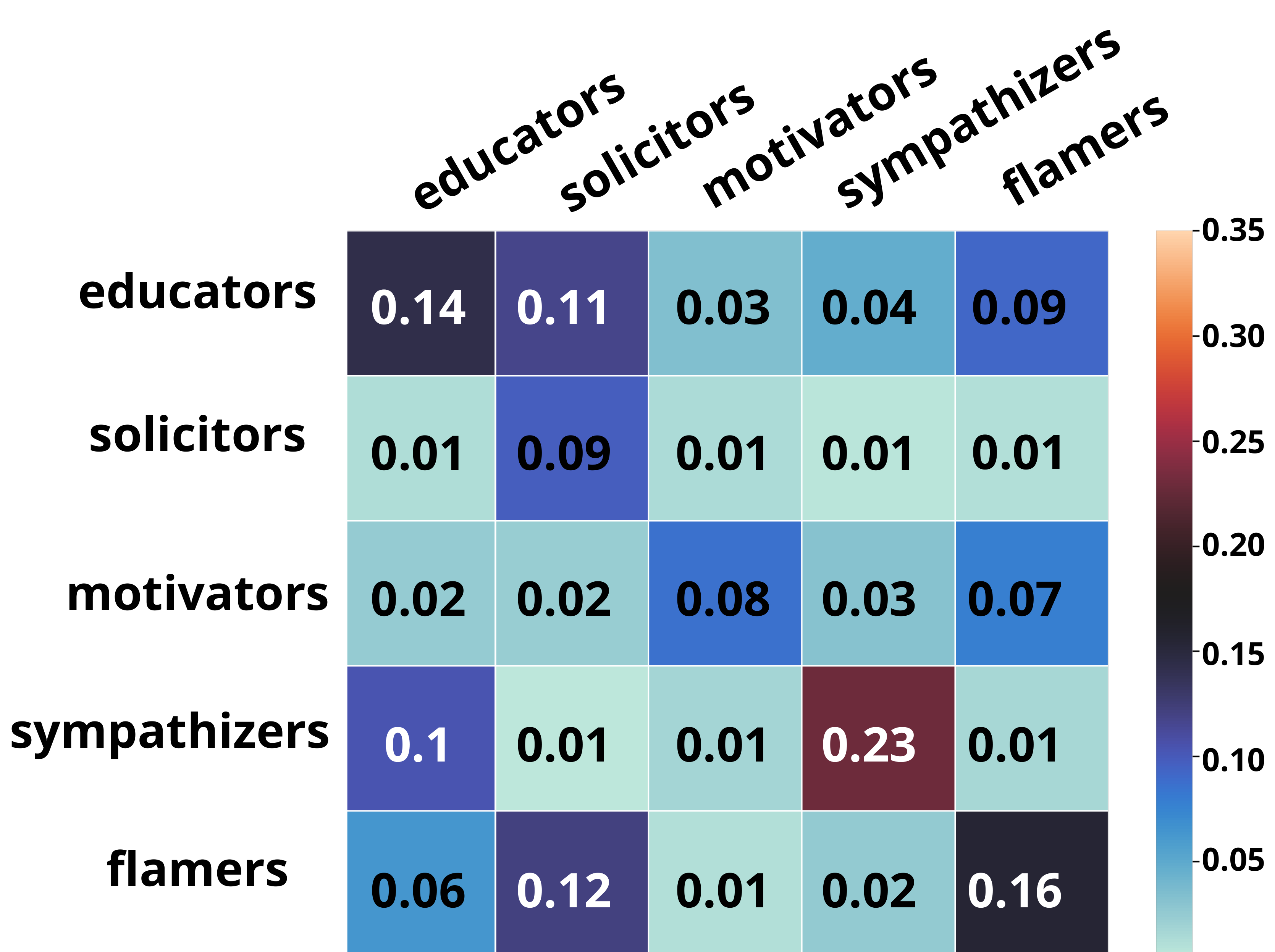}
        \caption{Fake news sources}
    \end{subfigure}%
    ~ 
    \begin{subfigure}[t]{0.4\textwidth}
        \centering
        \includegraphics[width=0.90\textwidth]{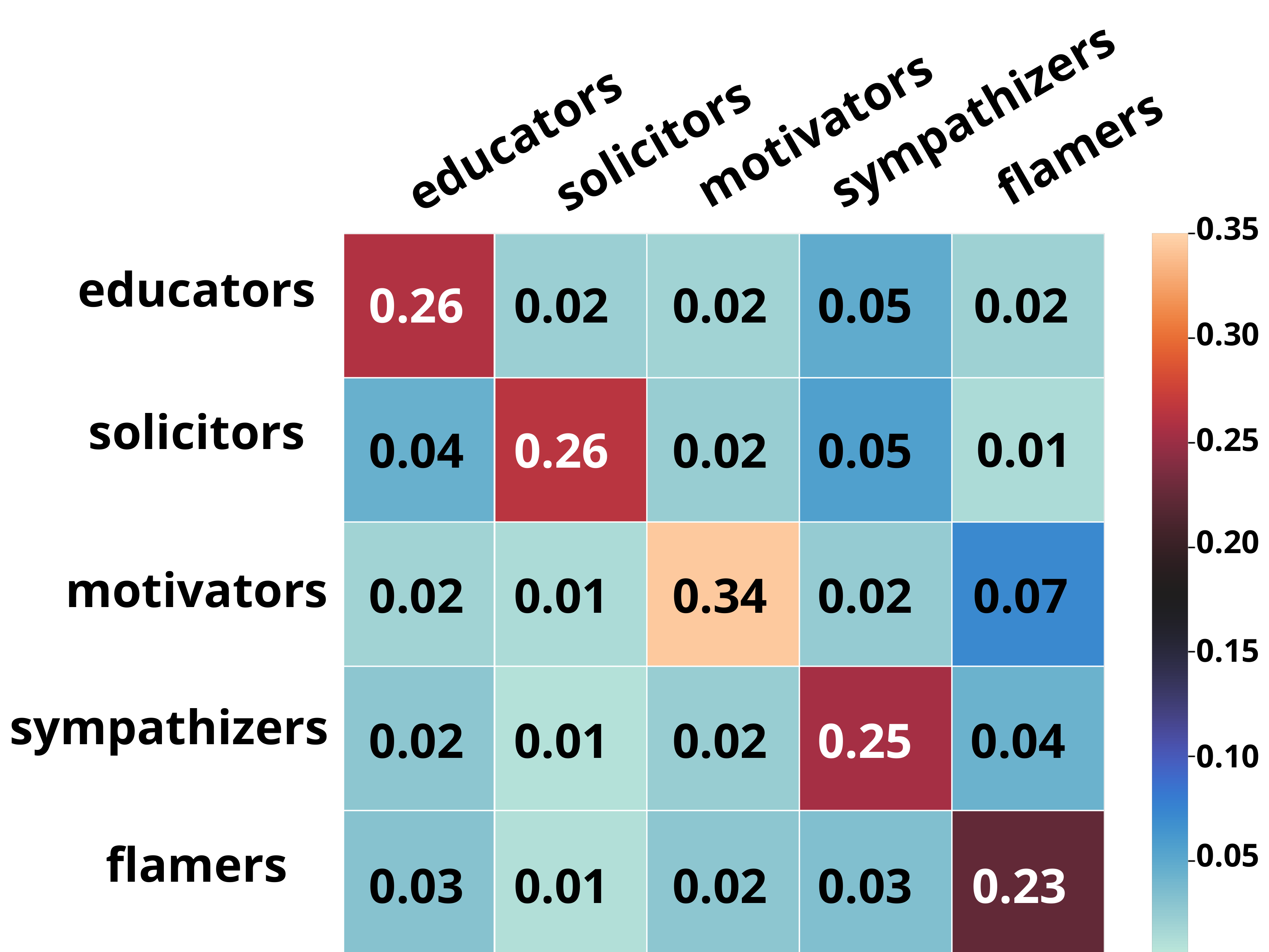}
        \caption{Conspiracy sources}
    \end{subfigure}

    \caption{Figure displaying the pairwise influence matrix for each type of source. The value of $W_{solicitors \rightarrow sympathizers} = 0.21$ in (a) indicates how influential \textit{solicitors} are in triggering extremist information sharing by \textit{sympathizers}. We have used divergent color palette to distinguish between the smaller non-diagonal values that are of most relevance while discussing the results.}
    \label{fig:influence}
\end{figure*}

Remember that we obtained weight matrix $W$ by fitting a Hawkes model for links from every source type. We report the weight matrices for four source types in Figure \ref{fig:influence}. Value in weight matrices, for example,  ($W_{\textit{solicitor} \rightarrow \textit{educator}} = 0.11$ in Figure \ref{fig:influence} (a)) indicates the strength with which  \textit{solicitors} may trigger information sharing by \textit{educators} account. For all source types, the self weights for roles  (for example, $W_{\textit{solicitor} \rightarrow \textit{solicitor}} = 0.35$ in Figure \ref{fig:influence} (a)) are highest. In other words, the main diagonal in all weight matrices in Figure \ref{fig:influence} have the highest values. This means that extremist accounts in one role are more likely to trigger information sharing by other extremist accounts in the same role. This may be because the the accounts in the same role also share large number of same links. Note that the values for weights reported in Figure \ref{fig:influence} are typical of influence values observed in information sharing on social media \cite{zannettou2017web}. 
%This means that a link posting event by an account assuming role $i$, strongly affects the same link being posted by other accounts assuming the same role. This may be because the Facebook pages/groups assuming the same role---accounts with similar behavioural characteristics in sharing extremist content---also post large number of same links. 

\vspace{3pt}
\noindent \textbf{Extremist Domains: } In all, we had listed 289 extremist website domains. Here, we analyze how important every role is in posting  links from extremist domains in Facebook. Figure \ref{fig:influence}(a) displays pairwise influence of roles on each other. For example, the value of $W_{solicitors \rightarrow sympathizers} = 0.21$ indicates how likely the link posting  by \textit{solicitors} will trigger the link posting by \textit{sympathizers}. From RQ1 analysis, we  know that \textit{solicitors} and \textit{educators} post large proportion of links from extremist websites. However, their higher influence on other roles indicates that they also trigger the spread of extremist content by other roles. Figure \ref{fig:educators_example} in Appendix displays a post by an \textit{educator} account citing a link from extremist domain discussing white identity. The link cited in this post is highly education in that it describes the constructs of white identity and implores the reader to not engage in interracial relationships.  Surprisingly, \textit{flamers} also have higher influence compared to \textit{sympathizers} ($0.13$) and \textit{motivators} (0.11). Overall, these results indicate that \textit{solicitors}, \textit{educators}, and \textit{flamers} are most influential in spreading links from extremist sources.  

\vspace{3pt}
\noindent \textbf{Biased Domains: } As per OpenSources, biased news category contains sources that hold a specific point of view and may rely on propaganda, decontextualizing information and opinions distorted as facts \footnote{\url{https://douglasducote.com/wp-content/uploads/2019/05/OpenSources.pdf}}. %For example,  OpenSources.co lists biased sources such as {\small \tt 100percentfedup.com	}---website that selects stories with extreme right wing bias \footnote{\url{https://mediabiasfactcheck.com/100-percent-fed-up/}} and {\small \tt dailywire.com}---media source strongly biased toward conservative causes and/or political affiliation \footnote{\url{https://mediabiasfactcheck.com/the-daily-wire/}}. 
For example,  OpenSources.co lists {\small \tt 100percentfedup.com} and {\small \tt dailywire.com} as biased sources. While {\small \tt 100percentfedup.com} selects stories with extreme right wing bias \footnote{\url{https://mediabiasfactcheck.com/100-percent-fed-up/}}, {\small \tt dailywire.com} is strongly biased toward conservative causes and/or political affiliation \footnote{\url{https://mediabiasfactcheck.com/the-daily-wire/}}.
Overall, the influence of all roles is lower while sharing links from the biased sources. However, in comparison to other roles, \textit{motivators} are more influential in triggering biased information posting by other roles ($W_{motivators \rightarrow educators} = 0.07$, $W_{motivators \rightarrow sympathizers} = 0.05$)

\vspace{3pt}
\noindent \textbf{Fake News Domains: } According to OpenSources, fake news sources  entirely fabricate information or grossly distort actual news reports. In the qualitative evaluation with experts, we had observed that \textit{flamers} post links that are flagged as fake/misinformative by Facebook fact checkers. Through weight matrix in Figure \ref{fig:influence} (c) we quantitatively observe that \textit{flamers} have higher influence on \textit{solicitors} ($0.12$) in spreading links from fake news sources. Figure \ref{fig:flamers_example} in Appendix displays how \textit{flamers} might be posting fake news to spread hate and outrage. Interestingly, \textit{educators} are also influential in spreading fake news to \textit{solicitors} ($W_{educators \rightarrow solicitors} = 0.11$).  However, it should be noted that fake news sources have lowest number of links and link posting events amongst all source types (Table \ref{tab:event_table}). Overall, based on the weight matrix, \textit{flamers} and \textit{educators} are most influential in spreading links from fake news sources.

\vspace{3pt}
\noindent \textbf{Conspiratorial Domains: } \textit{Sympathizers} are most susceptible to conspiracy information from \textit{solicitors} ($0.05$) and \textit{educators} (0.05).  Similarly, \textit{flamers} are susceptible to conspiratorial information sharing by \textit{motivators}($0.07$).  
%Figure \ref{fig:cpnspiracy_example} in Appendix displays posts by \textit{motivators} and \textit{educators} using conspiracy sources to motivate action against the targeted groups. 
The weights in the Figure \ref{fig:influence} (d) indicate that \textit{solicitors}, \textit{educators} and \textit{motivators} are most influential in disseminating links from conspiratorial sources. 
  
% Please add the following required packages to your 

\section{\textbf{Discussion and Implications}}
In this work we identify 5 functional roles in online extremist movements and investigate which roles are influential in spreading links from extremist, biased news, fake news and conspiracy sources. %To the best of our knowledge, we present the first work to use theoretically driven features to model participation roles in Facebook extremist communities. 
Our results can offer insights into how participatory activism advances extremist movements,  how various roles are located on the pathways to deeper engagement into extremism, and what could be the possible effects of interventions in countering online extremism undertaken by these roles. %Our work also has implications towards extending theories pertaining to social movement participation and in informing our understanding of the effects of interventions in countering online extremism undertaken by various roles. 

\subsection{\textbf{Online Extremist Movements and Participatory Activism}}
%Here, we reflect on how participatory activism might advance extremist movements and pose a threat to overall well-being on social media. 
Scholars have attributed the advancement of social movements to the successful distribution of resources through its participants \cite{MccarthyResourceTheory}. We observe that through participatory activism, extremist accounts, in various roles, adequately use social media to spread various types of information resources. For example, \textit{educators} and \textit{solicitors} dedicate a large proportion of their Facebook activity to distributing extremist content for educating and soliciting the readers into extremist movements. Moreover, the results of our third research question suggest that they also influence other roles in spreading information from extremist websites. In sum, by disseminating information through their Facebook accounts, mass educating the readers about their agenda, and soliciting funds and participation in the movements, \textit{educators} and \textit{solicitors} are creating human and material resources \cite{MccarthyResourceTheory}. Moreover, by prominently sharing misinformative content and using toxic language, \textit{flamers} may be raising emotional resources that create opportunities for public outrage and eventually, collective action \cite{van2017individuals} to advance the hateful agendas of their extremists movements. %Such human, material, and emotional resources can then be converted into collective action towards the advancement of the movement \cite{MccarthyResourceTheory}. 
This distributed system of online information mobilization---distribution of various information resources through various roles online---can be compared to the democratization process in participatory activism \cite{krona20195}. The digital democratization process specifically consists of more equitable sharing of informational resources amongst the participants \cite{carroll2006democratic}. For example, by influencing other roles in sharing links from extremist domains, \textit{solicitors} and \textit{educators} are also \emph{empowering} others with those extremist information. Take for example, the Facebook page of Alliance Defending Freedom (ADF)---an \textit{educator} account in our study, representing a leading anti-LGBTQ organization \cite{adfsplc} that the Southern Poverty Law Center has tracked for decades. This organization started with a small group of christian leaders advocating for discredited practice of conversion therapy, criminalization of LGBTQ sexual acts and opposition to the transgender rights. ADF started with 84K Facebook page likes in 2012. Today, ADF's Facebook page has over 1.7 million page likes and over 1.6 million followers. Our dataset also revealed that over 1K other Facebook groups and pages have already shared content linking to ADF's website. With 1.6 million direct followers on the ADF's official Facebook page and an additional indirect exposure to users through shares on other Facebook pages, the material created by ADF %(through the posts made by other 1K pages containing links from ADF website), ADF 
is able to reach a vast audience. %and thus, build a network of supporters. 
This may suggest that, through participatory activism, the picture of online extremism has now shifted from a few selected radical websites and accounts to a spectrum of allies with access to extremist information and the affordances to share it with the mass.

\subsection{\textbf{Theoretical Implications: Parallels Between Theoretical and Online Roles}}
Some of our roles correspond to the categories of participants identified in theoretical research based on physical protest events and social movements. For example, the \textit{educators}---accounts that primarily focus on distributing links from extremist domains---may correspond to ``constituents'' as described by McCarthy and Zald \cite{MccarthyResourceTheory}; ``constituents'' are primary distributors of resources. Similarly, our \textit{solicitors}---who actively solicit participation via donations and gatherings---may correspond to ``beneficiary constituents'' \cite{MccarthyResourceTheory} who stand to gain from the success, funds, and connections emerging from the movement. The \textit{sympathizers} category may be similar to ``bystanders''---a group of third-party participants, as defined by Turner et. al. \cite{turner1969public}, who might acknowledge grievances related to the issues of social movement and take a sympathetic stand. 
%Understanding the similarities of our data-driven roles with the traditional taxonomy of the movement participation allows us to comment on the dynamics of the social movements participants. For example, constituents mobilize resource in order to convert other participants into constituents \cite{MccarthyResourceTheory}. Meaning, that the \textit{solicitors} and \textit{educators} in our dataset might be at the core of extremist movements actively trying to proselytize accounts from other roles. 
However, the \textit{motivator} role does not resemble any of the theoretically described categories. We believe that the \textit{motivator} role is specifically relevant in the online setting for relaying positive news and wins related to the extremist causes. Additionally, \textit{flamers} also do not correspond to any theoretical roles. Through our data driven methods, we are able to surface these new roles that characterize online participation in extremist social movements. We believe that our framework for identifying roles based on the characteristics of participation can be extended to other social movements as well. For example, studies investigating positive social movements, such as environmentalist movement, can adopt our methodologies and identify roles and their influence in popularizing environmentalist agenda. 

% studies investigating positive social movements can also be abstracted into our theoretically motivated role identification methodology. 

% our methods of role identification can be abtracted 

% while researchers report negative emotions as one of the \emph{many} drives in participation\cite{van2017individuals}, we find a whole group of extremist accounts---\textit{flamers}---using distinctively higher words related to anger and injustice while sharing links from the extremist websites. 

% This indicates that our methods and results have implications in surfacing new roles---\textit{motivators} and \textit{flamers}---that are specifically catered to online setting. Moreover, our methods can also be used to identify roles in positive social movements such as racial justice movement or environmentalist movements and consequently extending the theories of online movement participation. 
%extending the theoretically identified roles to online setting which may give rise to new roles. 

\subsection{\textbf{Trajectories of Extremist Movement Participation}}
Klandermans et. al. proposed a trajectory of social movement participation comprising four steps \cite{klandermans1984mobilization,klandermans1987potentials}.  First, people must sympathize with the ideals and goals of the movement, thus turning into potential targets for mobilization. Next, they must be targeted by core members' mobilization attempts. Next, they must develop motivation to participate in the movement and finally overcome possible barriers and engage in collective action \cite{sturmer2003dual}. Through our analysis, we identify a group of accounts played the role of \textit{sympathizers}---pages/groups expressing sympathy towards the extremist causes without getting heavily involved. 36.4\% of the accounts in our dataset are \textit{sympathizers}. Based on Klandermans's models, \textit{sympathizers} can also be viewed as the biggest potential group of supporters for the extremist movements. Interestingly, we also see that \textit{educators}, \textit{solicitors}, and \textit{flamers} have high influence on \textit{sympathizers} in spreading extremist content (Figure \ref{fig:influence} (a)). This is the second step in the Klandermans's model, whereby \textit{sympathizers} are targeted for mobilization. In other words, sympathizers may lie on the first two steps of the Klandermans's trajectories of participation. The third step consists of participants who have developed motivation for  participating. The \textit{flamer} and \textit{motivator} roles are primarily driven by motivating factors such as anger, injustice, and the sense of achievement. Hence, \textit{flamers} and \textit{motivators} might be on the third step of  Klandermans's trajectory. Finally, we believe that \textit{solicitors} and \textit{educators} are on the last step of the participation trajectory as they actively try to educate and proselytize others through collective action. In summary, based on Klandermans's comprising, \textit{developing sympathy} $\rightarrow$ \textit{getting targeted by mobilization} $\rightarrow$ \textit{developing motivation} $\rightarrow$ \textit{collective action}, is equivalent to the following role transitions: \textit{sympathizers} $\rightarrow$ (\textit{flamers} or \textit{motivators}) $\rightarrow$ (\textit{educators} or \textit{solicitors}), which together represents a trajectory of deeper engagement into the extremist movements online. 
Can targeted interventions for counter-extremism stop users from getting induced into the deep trenches of extremist movements? Below, we discuss which roles could potentially benefit from interventions designed for countering online extremism. 

\subsection{\textbf{Practical Implications: Interventions for Online Extremism Engagement}}
%In this work, we identified five roles in online extremist movements such as white supremacy and anti-LGBTQ. 
While this study focuses on identifying roles, it can also inform the design of interventions for countering extremism. Our results suggest that while accounts core to the extremist movements---\textit{educators} and \textit{solicitors}---tend to retain their roles, others are more likely to transition to different roles. For example, \textit{flamers} and \textit{motivators}  become \textit{sympathizers} with high probability. \textit{Flamers}, \textit{motivators} and \textit{sympathizers} also show more sporadic engagement with sharing extremist links compared to the \textit{educators} and \textit{solicitors}. A study by Siegel and Badaan revealed that targeted interventions against hate speech, such as sanctions on hateful messages,  leads users to tweet less hateful content, especially if the individuals are less engaged with the hate speech in the first place \cite{siegel2020no2sectarianism}. On the other hand,  accounts that frequently see or produce hostile language are less likely to get deterred by sanctions and may even express backlash  \cite{siegel2020no2sectarianism}. Other researchers also report that rather than conforming to the community norms upon receiving sanctions, the producers of hostile content are more likely to move to other platforms \cite{newell2016user} or find creative ways of continuing their hate speech \cite{chancellor2016thyghgapp}. For example, recall that the Facebook page, \textit{Pissed off White Americans}, described in the Introduction, shared videos that are now banned on YouTube. However, they still made the extremist videos available to the readers by hosting them on bitchute.com which has been described as the ``hotbed for violence and hate'' \cite{BitChute78online}. 
Considering this, our results suggest that \textit{flamers}, \textit{motivators} and \textit{sympathizers}---accounts infrequently exposed to extremist content---might benefit most from targeted interventions designed to counter extremism. On the contrary, \textit{educators} and \textit{solicitors} may retaliate or relocate to alternate platforms in response to an intervention.

\section{\textbf{Limitations and Future Directions}}
Our work has some limitations which also open up promising future directions. First, our dataset contains only US-based extremist websites and most hold far-right political ideology.  However, this skew towards far-right might not correctly represent the political scenarios from other countries. For example, unlike the U.S., Germany has observed increased political violence from both, far-left and far-right ideological groups \mbox{\cite{GermanyE53online}} and is known to have a history of violence from both ends of the political spectrum \mbox{\cite{jungkunz2019towards}}. Hence, while applying our study results in the context of other countries, researchers need to be cautious about the distribution of extremist ideologies in our dataset. 
%\hl{First, our dataset largely contains websites belonging to the extremist groups on the far-right of the U.S political spectrum. Future researchers can use our methodologies to study information ecosystem of the far-left extremist groups in the U.S and compare participatory activism across the two ends of U.S political spectrum. Future researchers could add to this by including  far-left extremist groups and also compare participatory activism across the political spectrum. } 
Next, we compiled extremist accounts based on their sharing behavior, specifically the number of unique links they shared from known SPLC-designated extremist websites. While this is a common methodological choice made while choosing users/accounts for studying social media activity, a stricter selection criteria can be beneficial. For example, in addition to the frequency of the extremist link posts, extremist accounts can be selected based on the the topics discussed in the posts. Moreover, while our data collection spans the activity of extremist accounts in 2018 and 2019, it was collected post-hoc, in May 2020. Once a Facebook page/group is banned, its data is no longer accessible through CrowdTangle or any official Facebook API. Thus, considering the recent Facebook bans on the white nationalist accounts, our study doesn't contain the data for the extremist accounts that were active in 2018-2019 but got banned before May 2020. In future, researchers can collect data in real time and extend our study to understand what roles among the extremist accounts face highest moderation.
Additionally, our RQ2 analysis does not account for potential cohort effect. Specifically, in RQ2, we measure role transition for extremist accounts in various time periods, across 2018 and 2019. While we normalize the features to control for overall trends happening on social media, external events happening between 2018 and 2019 (e.g.,  a sudden popularity of an extremist content or change in the language surrounding issues of extremism) could affect different users differently introducing cohort effects. In future, researchers can adapt our methods and extend them to controlled experiments so as to mitigate period and cohort effects.
Moreover, we observe the influence of various roles in spreading information on Facebook based on the link posting activity in time. Naturally, this approach does not account for the various modes of information, such as images, memes, screenshots, videos, web articles etc. and how that might effect the role dynamics. Measuring influence for various modes of information can lead to interesting findings such as, which roles are influential in propagating memes related to extremist issues. Moreover, our influence measurement is based on the temporal characteristics of information flow. A deeper qualitative analysis of link chains and the comments on the links could help establish causal relationship of influence between various roles. 
Finally, our results reveal the information ecosystem of extremist movements on just one platform---Facebook. The problem of extremism is also evident on other platforms such as Twitter and Gab. Indeed, researchers have shown that the extremist movements leverage different social media platforms towards different goals such as radicalization and mass education \cite{Phadkecross}. 
We encourage future researchers to extend our methods to model roles in extremist movements on other platforms or even, across the platforms.

\section{\textbf{Conclusion}}
In this work we analyze the online ecosystem of extremist movements through the lens of participatory activism. Specifically, we identify five social roles in online extremist movements: \textit{educators} \textit{solicitors}, \textit{flamers}, \textit{motivators} and \textit{sympathizers}. We also investigate the role dynamics and influence exerted by roles in spreading links from extremist, fake news, biased, and conspiratorial sources. 
We not only find that the roles core to the extremist movement (\textit{educators} and \textit{solicitors}) are more stable but that they also have higher influence on other roles in spreading extremist content. Our findings offer a perspective on how participatory activism might be advancing extremist movements and how various roles may be targeted for mobilization. Our results also have implications in extending theories of social movement participation and understanding the effectiveness of online counter-extremism strategies. 
%Additionally, we identify roles based on the mixed method approach---clustering based on the features followed by qualitative evaluation by experts. While the qualitative evaluation provides us with an independet assessment of clusters, it can also be extended to create 

%might be more successful than confront \textit{educators} or \textit{solicitors} who are more committed and consistent members. 

\section{Acknowledgments}
This paper would not be possible without the expert feedback by the entire Pennebaker Language Lab directed by Dr. James Pennebaker at University of Texas Austin. We also want to acknowledge the valuable feedback from the members of Social Computing Lab at Virginia Tech and University of Washington, Seattle. This project was partially funded by the Minerva Research Initiative.

\bibliographystyle{ACM-Reference-Format}
\bibliography{sample-base,references,local-ref,acmart,borrowed,ref1}

\appendix

\newpage
\section{\textbf{APPENDIX}}

\subsection{\textbf{Cluster Analysis with Various Link Thresholds}}

In RQ1, we select extremist accounts that post atleast 10 unique extremist links in the analysis time window. While we select this threshold due to the 95th percentile of the extremist link posting distributions across all the accounts, here we present analysis for different thresholds. Specifically, we consider accounts that would have been added to the dataset if the threshold was \textgreater{8}, \textgreater{6} and \textgreater{4} extremist links posted. We find that the majority of the new accounts would have been added to the \textit{sympathizers} category. This is not surprising as \textit{sympathizers} have lowest proportion of extremist links compared to other roles. We also experimented with keeping all the accounts; i.e, accounts that post atleast 1 extremist link in two years. Keeping all accounts added great amount of noise in the feature calculation process. This may be because many defining clustering features (opinions, solicitations, popularity, proportion, trends) are calculated explicitly on the text and frequency of the extremist links and the accounts with only one or two extremist links may not have any words or phrases related to solicitation, opinions or any reactions on the posts. Overall, our post-hoc analysis suggests that even by lowering the threshold to >4 extremist links per account, the constitution of the educators or solicitors---roles core to the extremist movements---does not change significantly.

% \usepackage{booktabs}
% \usepackage{multirow}
% \usepackage{graphicx}
\begin{table*}[]
\centering
\resizebox{\textwidth}{!}{%
\begin{tabular}{@{}ccccccc@{}}
\toprule
\multirow{2}{*}{\textbf{\begin{tabular}[c]{@{}c@{}}threshold for\\ extremist links\end{tabular}}} &
  \multirow{2}{*}{\textbf{\begin{tabular}[c]{@{}c@{}}accounts added\\ because of the new threshold\end{tabular}}} &
  \multicolumn{5}{c}{\textbf{clustering of additional accounts with original cluster centers}} \\ \cmidrule(l){3-7} 
                     &      & \textbf{sympathizers} & \textbf{motivators} & \textbf{flamers} & \textbf{educators} & \textbf{solicitors} \\
\textgreater 8 links & 536  & 86\%                  & 7\%                 & 4\%              & 2\%                & 1\%                 \\
\textgreater 6 links & 1023 & 81\%                  & 5\%                 & 10\%             & 2\%                & 2\%                 \\
\textgreater 4 links & 9278 & 75\%                  & 12\%                & 9\%              & 3\%                & 1\%                 \\ \bottomrule
\end{tabular}%
}
\caption{Table showing the number of new accounts that could be added in the dataset with various thresholds for the number of extremist links. Our analysis is based on the threshold of \textgreater{}10 extremist links per account. The table also describes what percent of the new accounts fall across various roles using the cluster centers obtained in RQ1. The majority of the new accounts fall under the sympathizer role.}
\label{tab:threshold_check}
\end{table*}

\subsection{\textbf{Robustness Tests for Clustering}}
\label{sec:cluster_robust}

\subsubsection{\textbf{Feature Correlation}}
\begin{figure*}[t!]
    \centering
    \begin{subfigure}{0.67\textwidth}
        \centering
        \includegraphics[width=0.99\textwidth]{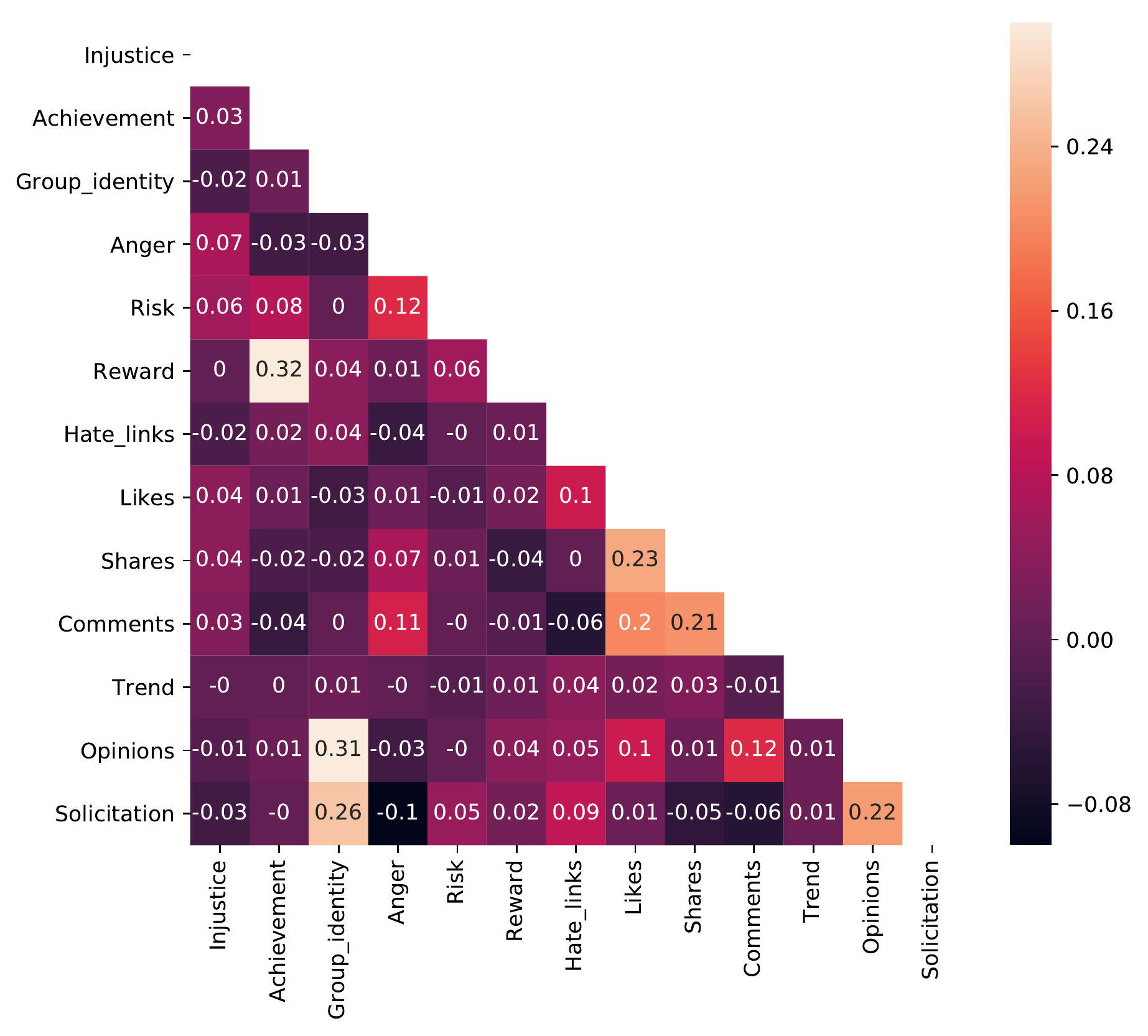}
        \caption{}
    \end{subfigure}%
    ~ 
    \begin{subfigure}{0.20\textwidth}
        \centering
        \resizebox{\textwidth}{!}{%
        \begin{tabular}{@{}rr@{}}
        \toprule
        \textbf{Features} & \multicolumn{1}{l}{\textbf{VIF}} \\ \midrule
        Injustice & 1.099729 \\
        Achievement & 1.882811 \\
        Group\_identity & 1.611063 \\
        Anger & 1.547119 \\
        Risk & 1.375777 \\
        Reward & 1.707319 \\
        Hate\_links & 1.161227 \\
        Likes & 1.946095 \\
        Shares & 1.444380 \\
        Comments & 1.356092 \\
        Trend & 1.003054 \\
        Opinions & 1.671187 \\
        Solicitation & 1.254916 \\ \bottomrule
        \end{tabular}%
        }
        %\caption{Jaccard sim}
        
        \caption{}
    \end{subfigure}

    \caption{(a) Figure showing pairwise Pearson correlation coefficients for all features used in the clustering. Most of the features have low correlation. Reward is moderately correlated with Achievement and Group identity is moderately correlated with Opinions. There are no high correlations. (b) Table displaying the Variance Inflation Factor (VIF) for all features. VIF is used to detect multi-colinearity. VIF exceeding 5 indicates high multi-colinearity \cite{sheather2009modern} between that feature and the others. All our features have VIF < 2 indicating low or no multi-colinearity. }
    \label{fig:correlation}
\end{figure*}

We first report the pairwise correlation values between different features used in clustering in Figure \mbox{\ref{fig:correlation}}. Most feature pairs have low correlation. Only pairs of reward, achievement and group identity, opinions are moderately correlated. We further test for multi-colinearity using Variance Inflation Factor (VIF). VIF values of $>5$ indicate high multi-colinearity \mbox{\cite{sheather2009modern}}. All of our features have a VIF $<2$ indicating low or no multi-colinearity

% \begin{figure*}[h]
%     \centering
%     \includegraphics[width=0.65\textwidth]{figures/correlations.pdf}
%     \caption{Figure showing pairwise Pearson correlation coefficients for all features used in the clustering. Most of the features have low correlation. Reward is moderately correlated with Achievement and Group identity is moderately correlated with Opinions. There are no high correlations. All features were mean centered before clustering.}
%     \label{fig:correlation}
% \end{figure*}

\subsubsection{\textbf{Finding Number of Roles}}
\begin{figure*}[t!]
    \centering
    \begin{subfigure}{0.53\textwidth}
        \centering
        \includegraphics[width=0.90\textwidth]{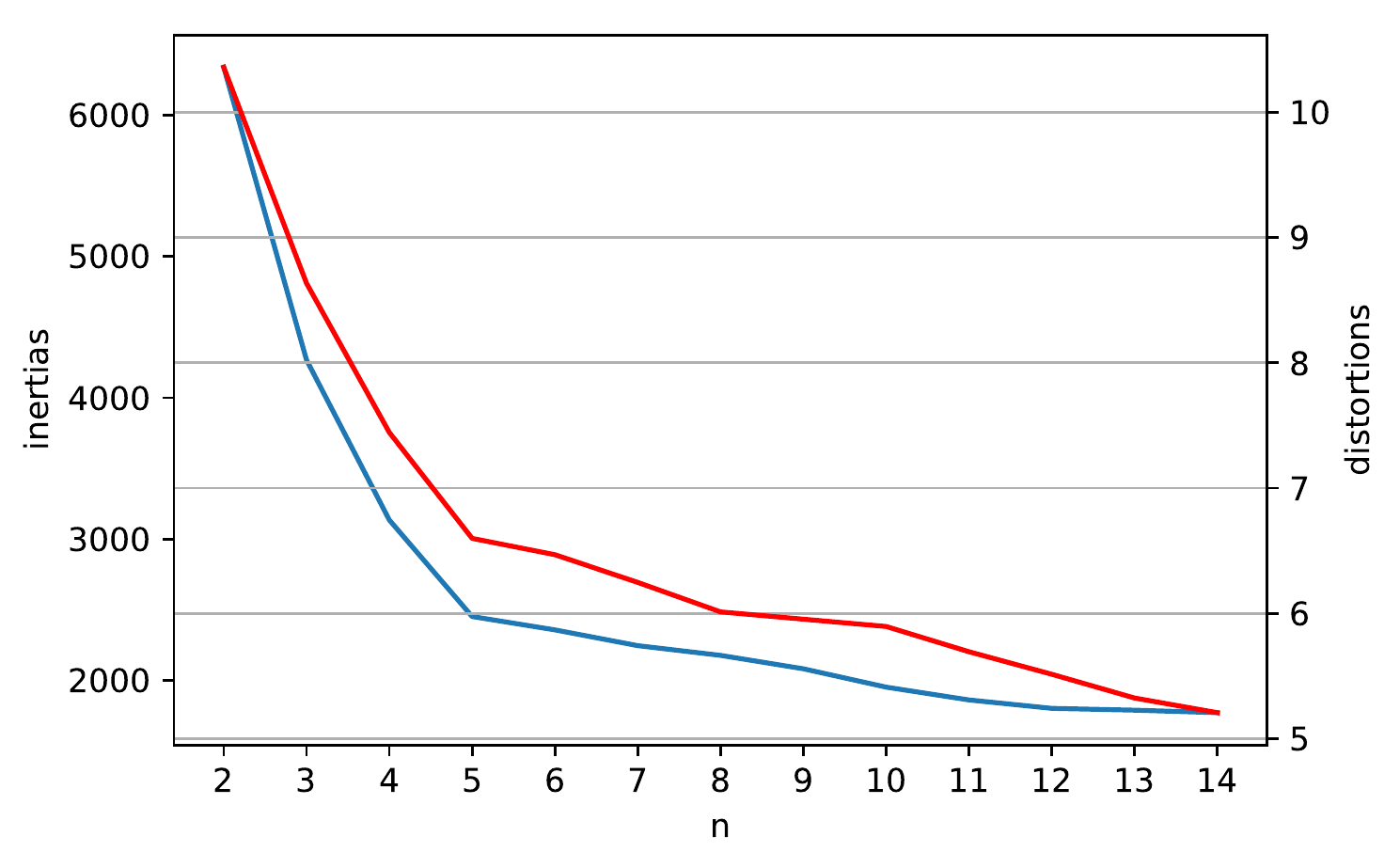}
        \caption{}
    \end{subfigure}%
    ~ 
    \begin{subfigure}{0.46\textwidth}
        \centering
        \resizebox{\textwidth}{!}{%
        \begin{tabular}{@{}cc@{}}
        \toprule
        \textbf{cluster} (role in KMeans) & \textbf{Jaccard(KMeans, Agglo)} \\ \midrule
        \textit{\textbf{A} (Solicitors)} & 0.87 \\
        \textit{\textbf{B} (Educators)} & 0.83 \\
        \textit{\textbf{C} (Flamers)} & 0.79 \\
        \textit{\textbf{D} (Motivators)} & 0.67 \\
        \textit{\textbf{E} (Sympathizers)} & 0.71 \\ \bottomrule
        \end{tabular}%
        }
        %\caption{Jaccard sim}
        
        \caption{}
    \end{subfigure}

    \caption{Figure presenting the elbow analysis using different algorithms. (a) Plots showing distortion and inertia across different number of clusters. Optimal number of clusters can be determined by the ``elbow'' of the curves. (b) Jaccard similarity scores between corresponding clusters of KMeans and Agglomerative clustering algorithms. }
    \label{fig:elbow}
\end{figure*}

In order to determine the number of roles, we performed elbow analysis with KMeans algorithm. We used two metrics---inertia and distortions---and plotted their values across different number of clusters (or roles). Distortions are calculated as the average of the squared distances from the cluster centers and inertia is the sum of the squared distance from the cluster centers. Figure \mbox{\ref{fig:elbow}} (a) displays inertia and distortion plots for the KMeans algorithm where elbow can be observed at $n=5$ indicating that 5 is the optimal number of roles. 
%To further validate the choice of number of roles, we calculated Akaike information criterion (AIC) and Bayesian information criterion (BIC) using Gaussian Mixture Model (GMM). Figure \ref{fig:elbow} (b) displays AIC and BIC plots showing elbow around $n=5$, confirming our choice for the number of roles. 

\subsubsection{\textbf{Cluster Assignments Overlap}}
We obtained role assignment for every extremist account using KMeans algorithm. As a further robustness test, we also compare the role assignments done by  KMeans with a different algorithm---Agglomerative clustering. We selected Agglomerative clustering as it is hierarchical clustering algorithm operating on a different clustering mechanism compared to KMeans.  
Note that since KMeans and Agglomerative clustering are inherently motivated by different working principles, it is harder to achieve perfect similarity between cluster assignments. However, we present Jaccard similarity scores for cluster assignments using KMeans and Agglomerative clustering to provide an estimate of overlap between the two algorithms. We fist cluster extremist accounts using KMeans algorithm and record cluster assignments. We then cluster the accounts using Agglomerative clustering and record the cluster assignments. For every cluster in the KMeans (eg., KMeans\_A), we find corresponding cluster in the Agglomerative clustering (eg., Agglo\_A) and calculate Jaccard similarity score using common extremist accounts in both clusters. Specifically, we calculate the number of common extremist accounts in Kmeans\_A and Agglo\_A and divide it by total number of unique accounts in Kmeans\_A and Agglo\_A combined. Figure \mbox{\ref{fig:elbow}} (b) provides the Jaccard similarity values for similar clusters obtained by two different algorithms. On an average the two algorithms have 0.77 similarity in cluster assignments.

\subsection{\textbf{Tables for Opinion and Solicitation Expressions}}

% \usepackage{booktabs}
% \usepackage{graphicx}
\begin{table}[h]
\centering
\scriptsize
\resizebox{0.8\textwidth}{!}{%
\begin{tabular}{@{}llll@{}}
\toprule
\textbf{Pronoun} & \textbf{POS} & \textbf{Variation} & \textbf{Examples} \\ \midrule
\begin{tabular}[c]{@{}l@{}}First person subjective\\ (I , we)\end{tabular} & \begin{tabular}[c]{@{}l@{}}LIWC cogproc verbs\\ (believe, think, infer, propose)\end{tabular} & \begin{tabular}[c]{@{}l@{}}With and without auxiliary verbs, \\ adverbs, negations\end{tabular} & \begin{tabular}[c]{@{}l@{}}I believe..\\ We don't think..\\ I really don't understand..\end{tabular} \\
\begin{tabular}[c]{@{}l@{}}First person possessive \\ (my, mine, our, ours)\end{tabular} & \begin{tabular}[c]{@{}l@{}}LIWC cogproc nouns\\ (opinion, understanding)\end{tabular} & With or without adjectives & \begin{tabular}[c]{@{}l@{}}My strong opinion…\\ Our shared understanding of the issue...\end{tabular} \\
\begin{tabular}[c]{@{}l@{}}First person subjective\\ (I , we)\end{tabular} & \begin{tabular}[c]{@{}l@{}}LIWC cogproc adjectives\\ (positive, confused, unclear, \\ hopeful)\end{tabular} & \begin{tabular}[c]{@{}l@{}}With and without auxiliary verbs, \\ adverbs, negations\end{tabular} & \begin{tabular}[c]{@{}l@{}}I am hopeful…\\ We are really confused…\\ I might not be supportive..\end{tabular} \\
\begin{tabular}[c]{@{}l@{}}Third person \\ (he, she it, they, theirs,\\  his, hers, its)\end{tabular} & \begin{tabular}[c]{@{}l@{}}Modal verbs\\ (should, must, can , could, \\ will, might)\end{tabular} & \begin{tabular}[c]{@{}l@{}}With ot without nouns, negations, \\ adverbs in the middle\end{tabular} & \begin{tabular}[c]{@{}l@{}}They should...\\ He should...\\ They definitely must...\\ His incomplete understanding must...\end{tabular} \\
Proper nouns & \begin{tabular}[c]{@{}l@{}}Modal verbs\\ (should, must, can , could, \\ will, might)\end{tabular} & \begin{tabular}[c]{@{}l@{}}With ot without nouns, negations, \\ adverbs in the middle\end{tabular} & \begin{tabular}[c]{@{}l@{}}Hillary must...\\ Trump should...\\ CAIR actually can...\end{tabular} \\ \bottomrule
\end{tabular}%
}
\caption{Table listing phrase patterns used to extract opinions. We pair various LIWC cognitive processing (cogproc) nouns, verbs and adjectives with pronouns. The last column also lists examples for each pattern.}
\label{tab:op_pat}
\end{table}

% \usepackage{booktabs}
% \usepackage{graphicx}
% \usepackage[normalem]{ulem}
\begin{table}[h]
\centering
\resizebox{0.7\textwidth}{!}{%
\begin{tabular}{@{}llll@{}}
\toprule
\textbf{Pronouns/POS} & \textbf{Pronouns/POS} & \textbf{Variation} & \textbf{Examples} \\ \midrule
please & \begin{tabular}[c]{@{}l@{}}LIWC social verbs\\ (donate, call , register)\end{tabular} & \begin{tabular}[c]{@{}l@{}}With and without \\ auxiliary verbs, \\ adverbs\end{tabular} & \begin{tabular}[c]{@{}l@{}}Please donate…\\ Please consider registering..\end{tabular} \\
LIWC social verbs & LIWC social nouns & \begin{tabular}[c]{@{}l@{}}With or without article, \\ adjective, preposition\end{tabular} & \begin{tabular}[c]{@{}l@{}}Sign the petition...\\ Register for this beautiful event...\end{tabular} \\
LIWC social verbs & First person pronoun & \begin{tabular}[c]{@{}l@{}}With or without \\ prepositions\end{tabular} & \begin{tabular}[c]{@{}l@{}}Contact us..\\ Register with our...\end{tabular} \\
\begin{tabular}[c]{@{}l@{}}Second person\\ (you, yours)\end{tabular} & \begin{tabular}[c]{@{}l@{}}Modal verbs\\ (should, must, can , \\ could, will, might)\end{tabular} & \begin{tabular}[c]{@{}l@{}}With or without nouns, \\ negations, adverbs\end{tabular} & \begin{tabular}[c]{@{}l@{}}You should..\\ You really must..\\ You can..\end{tabular} \\
\begin{tabular}[c]{@{}l@{}}Second person\\ (you, yours)\end{tabular} & LIWC social noun & \begin{tabular}[c]{@{}l@{}}With or without negations, \\ adjectives  in the middle\end{tabular} & \begin{tabular}[c]{@{}l@{}}Your wonderful donation…\\ Your timely call..\end{tabular} \\
\begin{tabular}[c]{@{}l@{}}Verbs used in requests\\ (will, could, can, would)\end{tabular} & Second person &  & \begin{tabular}[c]{@{}l@{}}Will you..(sign this petition)?\\ Can you …(call your senator?)\end{tabular} \\ \bottomrule
\end{tabular}%
}
\caption{Table listing phrase patterns used to extract expressions of solicitation. We pair various LIWC social category nouns, verbs and adjectives with pronouns. The last column also lists examples for each pattern.}
\label{tab:sol_pat}
\end{table}

\subsection{\textbf{Analysis of the Domains removed in RQ3}}
For our RQ3 analysis, we removed all links in time period $T_{1}$ sourced from domains not present in the OpenSources dataset or our list of extremist websites. Specifically, we removed 24\% of the link posts. It is possible that some of the removed domains could contain extremist, fake news, biased or conspiratorial content. In order to estimate how many of the removed domains could potentially fall under extremist, fake news, biased or conspiratorial content categories, we report 100 most frequently shared domains from the removed links (See Table \mbox{\ref{tab:missdomains}}). For each of the 100 domains, we searched for the bias and credibility ratings using {\small \tt mediabiasfactcheck.com} and {\small \tt allsides.com}. In Table \mbox{\ref{tab:missdomains}} we highlight domains that have {\color{bias}{moderate to extreme political bias (left or right)}}, {\color{fake}{low or very low factuality rating}} and {\color{consp}{strong conspiracy or pseudoscience rating}}. 20 out of 100 domains are either strongly biased, have factuality or strong conspiratorial content.

\begin{table*}[h]
\tt
\small 
\centering
\resizebox{\textwidth}{!}{%
\begin{tabular}{ccccc}
\color{bias}{\textbf{foxnews.com}} &kirkreview.com & bible.com & cbsnews.com & jwatch.us \\
msn.com & newsweek.com & israelnationalnews.com & \color{consp}{\textbf{cbn.com}} & strangesounds.org \\
thehill.com & iheart.com & bloomberg.com & nationalpost.com & beecatholic.com \\
washingtonpost.com & \color{bias}{\textbf{hannity.com}} & themaven.net & haaretz.com & medium.com \\
cnn.com & \color{bias}{\textbf{freebeacon.com}} & wsj.com & \color{fake}{\textbf{rushlimbaugh.com}} & globalnews.ca \\
washingtontimes.com & \color{consp}{\textbf{endtimeheadlines.org}} & torontosun.com & foxbusiness.com & verelq.am \\
nytimes.com & \color{bias}{\textbf{gatestoneinstitute.org}} & politico.com & nydailynews.com & \color{bias}{\textbf{davidharrisjr.com}} \\
theguardian.com & independent.co.uk & businessinsider.com & enlacejudio.com & snopes.com \\
jpost.com & atheistrepublic.com & \color{bias}{\textbf{dmlnews.com}} & eventbrite.com & realclearpolitics.com \\
timesofisrael.com & whitehouse.gov & forbes.com & rightspeak.net & vox.com \\
nypost.com & bbc.com & wikipedia.org & time.com & whoobazoo.com \\
elderstatement.com & patheos.com & gofundme.com & mckoysnews.com & deborahhbateman.com \\
\color{fake}{\textbf{rt.com}} & dailymail.co.uk & \color{consp}{\textbf{globalskywatch.com}} & \color{bias}{\textbf{patriotbeat.com}} & ch7.io \\
\color{bias}{\textbf{townhall.com}} & fenixx.org & \color{bias}{\textbf{telegraph.co.uk}} & thetrumptimes.com & observador.pt \\
\color{bias}{\textbf{gellerreport.com}} & \color{bias}{\textbf{voiceofeurope.com}} & apple.news & aljazeera.com & algemeiner.com \\
freespeechtime.net & soundcloud.com & change.org & therebel.media & godaddy.com \\
huffingtonpost.com & latimes.com & npr.org & chicagotribune.com & aleteia.org \\
reuters.com & \color{bias}{\textbf{thefederalist.com}} & ynetnews.com & mirror.co.uk & cbslocal.com \\
usatoday.com & spencerfernando.com & palinfo.com & \color{fake}{\textbf{cnsnews.com}} & tapwires.com \\
nbcnews.com & cnbc.com & \color{bias}{\textbf{bizpacreview.com}} & msnbc.com & \color{fake}{\textbf{bitchute.com}}
\end{tabular}%
}
\caption{Table with 100 most frequent domains in URLs removed in the RQ3 analysis. We highlight domains that have {\color{bias}{moderate to extreme political bias (left or right)}}, {\color{fake}{low or very low factuality rating}} and {\color{consp}{strong conspiracy or pseudoscience rating}}. }
\label{tab:missdomains}
\end{table*}

\subsection{\textbf{Mathematical Formulation for Hawkes Process}}
\label{sec:math_hawkes}
Consider a series of link posting events divided into the time bins $\Delta$t
This series is a Hawkes process if the rate of each process has the parameterized form:

\begin{equation}
    \lambda_{t,k} = \lambda_{0,k} + \sum_{k^{'}=1}^{K} \sum_{t^{'}=1}^{t-1} s_{t^{'},k^{'}} \cdot h_{k^{'}\rightarrow k}[t - t^{'}]
\end{equation}

where $\lambda_{0,k}$ is the background rate, $s$ is the matrix of events generated from common link sharing and $h$ is an impulse response function describing the amplitude of influence that events on process (or role) $k^{'}$ have on the rate of process (or role) $k$. Guided by the descriptions in \cite{zannettou2017web,linderman2014discovering}, we can further decompose $h$ into the weight matrix $W$ and probability mass function $G$. We use the the weight matrix  $W$ of the shape $W^{K \times K}$ to infer pairwise influence of roles. 

\subsection{\textbf{Example Links from Various Source Types Posted by Influential Roles }}
\begin{figure*}[h]
    \centering
    \includegraphics[width=0.80\textwidth]{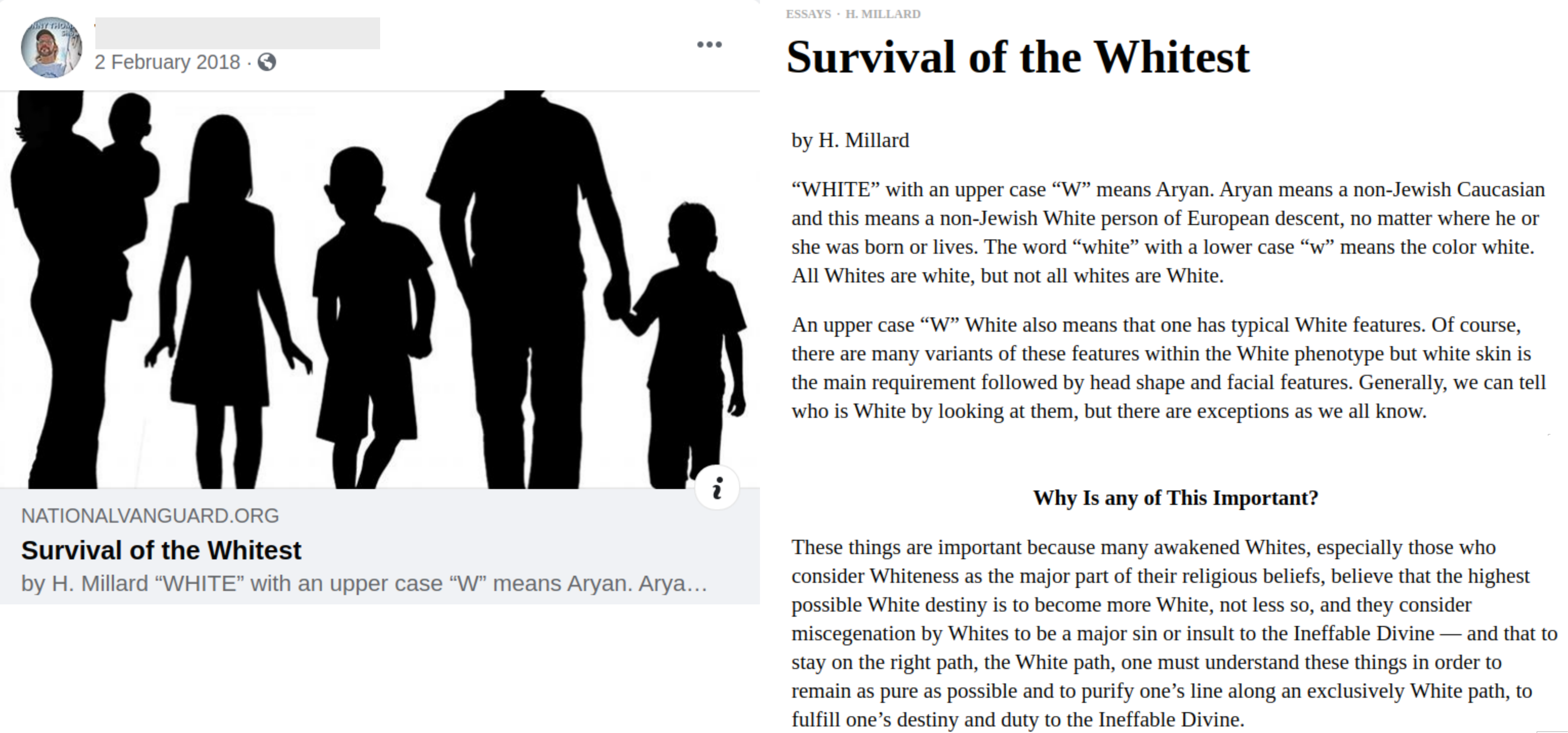}
    \caption{Example of link posted by \textit{educators} from an extremist domain about white identity. The content of the link is highly educational in that, it explains the constructs of the Aryan identity and emphasizes on avoiding interracial relations. }
    \label{fig:educators_example}
\end{figure*}

\begin{figure*}[h]
    \centering
    \includegraphics[width=0.80\textwidth]{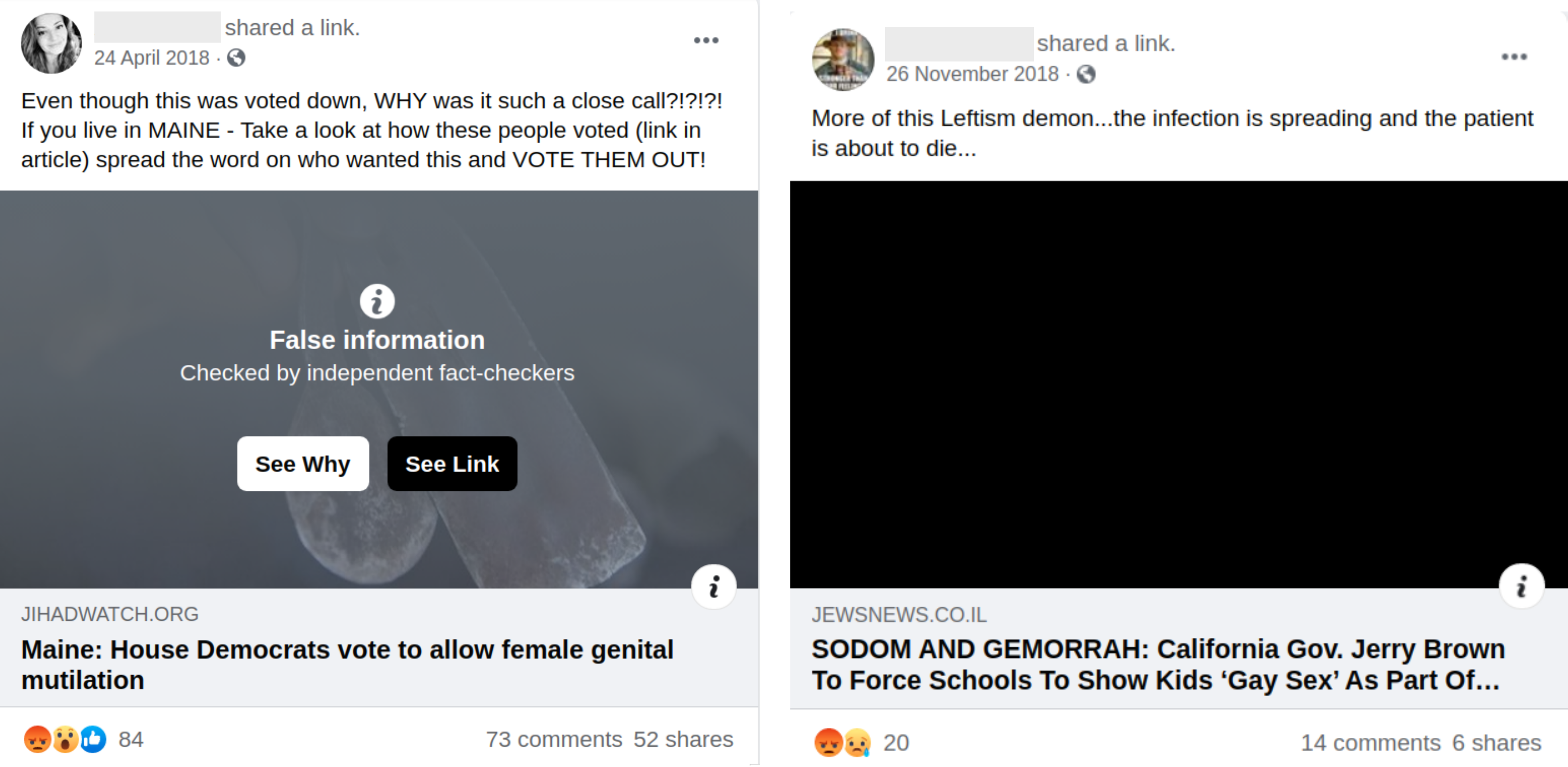}
    \caption{Example of links posted by \textit{flamers} from fake news domain. The first link presents false information to alarm and mobilize readers against the Democratic party. Second link uses false information alongwith fear based narratives to present anti-LGBTQ agenda.}
    \label{fig:flamers_example}
\end{figure*}

% \begin{figure*}[h]
%     \centering
%     \includegraphics[width=0.90\textwidth]{figures/conspiracy_example.pdf}
%     \caption{Examples of posts containing links from the conspiratorial domains.}
%     \label{fig:cpnspiracy_example}
% \end{figure*}

\end{document}